\documentclass{article}

\usepackage{microtype}
\usepackage{graphicx}
\usepackage{subfigure}
\usepackage{booktabs} 

\usepackage{hyperref}

\usepackage[accepted]{mlsys2022}

\mlsystitlerunning{An overview of the data-loader landscape: comparative performance analysis
}

\graphicspath{img}

\begin{document}

\twocolumn[
\mlsystitle{An overview of the data-loader landscape: comparative performance analysis
}



\mlsyssetsymbol{equal}{*}

\begin{mlsysauthorlist}
\mlsysauthor{Iason Ofeidis}{equal,yale}
\mlsysauthor{Diego Kiedanski}{equal,yale}
\mlsysauthor{Leandros Tassiulas}{yale}
\end{mlsysauthorlist}

\mlsysaffiliation{yale}{Department of Electrical Engineering, and Yale Institute for Network Science,
Yale University, New Haven, CT 06520, USA}

\mlsyscorrespondingauthor{Diego Kiedanski}{diego@kiedanski.com}

\mlsyskeywords{Machine Learning, MLSys, Dataloader, S3}

\vskip 0.3in

\begin{abstract}
Dataloaders, in charge of moving data from storage into GPUs while training machine learning models, might hold the key to drastically improving the performance of training jobs. Recent advances have shown promise not only by considerably decreasing training time but also by offering new features such as loading data from remote storage like S3. In this paper, we are the first to distinguish the dataloader as a separate component in the Deep Learning (DL) workflow and to outline its structure and features. Finally, we offer a comprehensive comparison of the different dataloading libraries available, their trade-offs in terms of functionality, usability, and performance and the insights derived from them. 
\end{abstract}
]



\printAffiliationsAndNotice{\mlsysEqualContribution}


\section{Introduction}\label{sec:introduction}

Training a (deep) machine learning model requires a dataset, a model, and the hardware, which for real problems involves one or more GPUs.

We are always interested in reducing the total computational time required to train a model. This is desirable for several reasons: lower costs, easier to iterate, and more accessible for smaller teams, among other things.

The relationship between the main components of an ML pipeline and the running time is often explicit: a larger dataset takes longer, a larger model takes longer, and a faster GPU reduces the total running time. One key piece in the puzzle that is often overlooked is the glue between all these parts: the dataloader. The dataloader is in charge of loading the data from its permanent storage (RAM, disk, or the network), applying the necessary transformations, and sending the transformed data to the appropriate device so the model can ingest it.

Most developers assume that the default dataloader in their respective machine learning framework (Pytorch, Tensorflow, Jax) is already optimized for their application and do not often rely on third-party data loaders. Interestingly enough, it has recently been shown that data loaders can be one of the more significant bottlenecks of ML pipelines\cite{mohan2020analyzing}. As a result, we have seen many new libraries and research projects dedicated to optimizing and improving the dataloader performance.

For example, FFCV \cite{leclerc2022ffcv}, a new open-source library developed by a research team at MIT, managed to train ImageNet in a fraction of the time it would take using the default PyTorch dataloader. Such gains can dramatically reduce the operational costs of companies and research teams that depend on infrastructure as a service (IaaS), such as Amazon Web Services (AWS) and Google Cloud Platform (GPC).

Another promising feature offered by dataloaders is the ability to load data stored remotely; for example, from an S3 bucket. This has many practical advantages: the time to set up the dataset locally is avoided, the required disk capacity on the computing machine is reduced, and the risk of team members using different versions of the same dataset is diminished. The natural drawback of having to stream the data while training is that, usually, network transfer speeds are slower than disk I/O, and, as a result, the model should take longer to train. Interestingly, we have observed that some libraries such as Hub \cite{2022ActiveloopHub} and Deep Lake \cite{deeplake2022}, achieve better performance over the network than the default Pytorch dataloader reading data locally for some scenarios. This is possible because the dataloader manages to pre-fetch the required data before the GPU needs it. We will offer a more extensive discussion in Section \ref{sec:discussion}.

Not all libraries support remote loading, and those that do, not necessarily integrate with the same remote storage services. Since the number of available libraries implementing dataloaders is growing, we set out to build a comprehensive benchmark to illuminate the current state-of-the-art, what problems seem to have been solved already and to discover the most promising areas for improvement in future research.

At this point, it should be mentioned that one particular difference from our experiments from other works such as \cite{abhishek2020quiver}, \cite{mohan2020analyzing} is that we focus on jobs that run on small to medium workstations with limited capacity (GPU, RAM, SSD). These are more likely to reflect the hardware available to most individuals and small teams in the industry, for whom the budget does not permit the usage of large-scale clusters.


\subsection{Contributions}

We can summarize our contributions as follows:

\begin{itemize}
    \item \textbf{Open-source Code}: We built an open source benchmark that compares the most popular data loading libraries in Pytorch\footnote{Github Repository: https://github.com/smartnets/dataloader-benchmarks}. The project will remain available to the community so new libraries and datasets can be added as interest in them increases. We also expect to update the numerical results obtained in these benchmarks following any major updates to any of the libraries benchmarked in this paper. 
    \item \textbf{Viability of Remote Training}: We show that it is possible to train a machine learning model using a data stream over a public internet connection under reasonable circumstances. In particular, we point out the impact of the computing serving the data. Our result is different than that of \cite{mohan2020analyzing} since we do not assume that the dataset is cached locally after the download.
    \item \textbf{Hyperparameter Optimization for Speed}: Traditional hyperparameter approaches aim to increase the overall accuracy of the model being trained. In this paper, we show how we can also optimize for speed (processed samples over time) as a proxy of total running time. This optimization is hardware-dependant, so it makes sense to perform it before long running jobs. This process should be at least an order of magnitude faster than equivalent time-to-acurracy metrics.
\end{itemize}

\section{Dataloaders}\label{sec:dataloaders}

In the process of training a Deep Learning model, the dataset needs to be read from memory and pre-processed, before it can be passed as input to the model. This operation requires loading the data into memory all at once. In most cases and, especially, with large datasets, a memory shortage arises due to the limited amount of it available in the system, which also deteriorates the system’s response time. This bottleneck is often remedied in deep learning libraries by using a so-called dataloader. This structure provides a way to iterate over the dataset by leveraging parallel processing, pre-fetching, batching, and other techniques to reduce data loading time and memory overhead as much as possible \cite{paszke2019pytorch}.

The main goal of a dataloader is to perform the actions responsible for transferring data samples from a storage location to the memory co-located with the processing units for training in order to form a batch of samples to be fed into the model. These actions are restrained by the storage system’s bandwidth and specifically its I/O performance. Thus, depending on the system’s hardware specifications, its filesystem serving it, and the throughput of the link with the computing units, it can have an immense influence on the total amount of time needed to complete the training.

The following specification of the dataloader component is mainly focused towards PyTorch (torch.DataLoader() \cite{pytorchwebsite}), while its TensorFlow counterpart (tf.Dataset()  \cite{tensorflow2015-whitepaper}), albeit not the same, bears great similarities.

When employing a dataloader, apart from providing the dataset for input, the user has the option to configure a number of hyperparameters, tailored to their needs and resources. A common one available in all dataloaders is the batch size, which, as mentioned before, defines the number of samples that will be used before updating the internal model parameters. This parameter is intrinsically linked with the concept of "mini-batch" in stochastic gradient descent \cite{hinton2012neural} and therefore, it is one of the first parameters that usually undergoes fine-tuning when one needs to achieve better training results.

Secondly, the user can define the sampling approach, which determines the strategy for drawing samples from the dataset and inserting them into a batch. This could include selecting samples based on specific criteria, as well as a probability distribution. In this step exists the option of shuffling, where the samples can be rearranged before every dataset iteration, with the goal usually being to improve the generalization of the training model. Another parameter is the collate/padding function, which essentially specifies the process of linking together all the individual samples inside a batch (think of stacking vectors into a tensor), in order to form a single element to be fed as input to the training model. Moreover, the dataloader can be configured to automatically store fetched data samples in pinned (page-locked) memory, thus enabling faster data transfer to CUDA-enabled devices.

Dataloaders come with a component called workers, whose purpose is to optimize this data-transferring process. Workers are defined as sub-processes responsible to carry out the data loading in an asynchronous fashion. When creating an instance of a dataloader, the user has the option to specify the number of workers that will be spawned and will be in control of this operation. If the number of workers is equal to zero, no sub-processes will be created, which, in turn, means that data fetching happens synchronously in the same process and, thus, the computing units (GPU) have to wait for the data loading to be completed \cite{pytorchwebsite}. Reversely, there will be generated sub-processes equal to the number of workers, which will prevent the blocking of computation code with data loading. This is accomplished by pre-fetching future batches in advance to be ready when needed.

\section{Experimental Setup}\label{sec:experimental}

Several libraries and datasets were selected to compare their features and performance. Even though an effort was made to be as comprehensible as possible, the field of data loading is constantly evolving and new libraries and releases are added every day. In this regard, we expect the following list to provide a good overview of the current capabilities of dataloaders, without necessarily claiming or finding the overall best (which would probably change from the time of writing to the time of publication). 
We make all the source code of the experiments available to the public and expect results to be fully reproducible.

\begin{figure}
    \centering
    \includegraphics{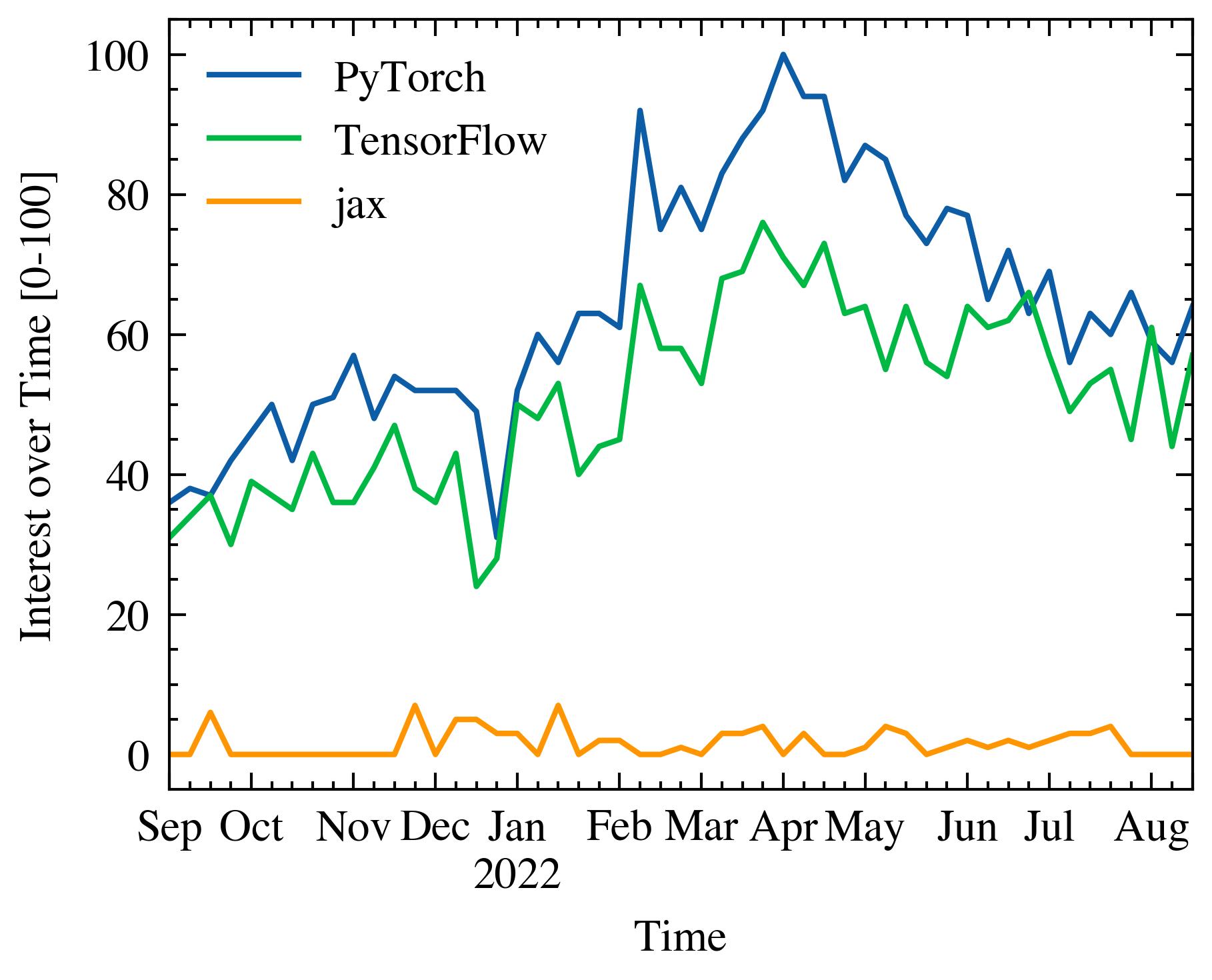}
    \caption{Popularity of different ML frameworks in the last 12 months}
    \label{fig:popularity}
\end{figure}

We selected seven libraries to perform our experiments: PyTorch \cite{paszke2019pytorch}, Torchdata \cite{TorchData}, Hub \cite{2022ActiveloopHub}, FFCV \cite{leclerc2022ffcv}, Webdatasets \cite{Webdataset}, Squirrel \cite{2022squirrelcore}, and Deep Lake \cite{deeplake2022}. One interesting thing that we discovered is that not all libraries support the same features. For example, we could not run FFCV with a dataset hosted in an S3 bucket to perform our remote experiments. As we mentioned in Section \ref{sec:introduction}, we run all our experiments in PyTorch. We considered reproducing the experiments in other popular machine learning frameworks but we decided against the idea since the second candidate would have been Tensorflow, but there are rumors that Google is moving away from it in favor of JAX. Figure \ref{fig:popularity} depicts the popularity of different ML frameworks in the last 12 months.

\subsection{Datasets}

Regarding the datasets, we initially opted for two classical datasets to support two different learning tasks: CIFAR-10 \cite{krizhevsky2009learning} for image classification, and CoCo \cite{lin2014microsoft} for object detection. While performing some experiments, we observed strange behavior (libraries performing better than expected) that could be explained by CIFAR-10 fitting into memory\footnote{This is often something desirable and expected in some of the reviews literature, but we found it not to be the case in several practical applications involving small teams and in-house workstations.}. For this reason, we built a third dataset named RANDOM, consisting of randomly generated colour images of size 256 by 256 pixels and a random class out of 20. This third dataset contains 45000 images for training, 5000 for validation, and 500 for testing, and it is considerably larger than CIFAR-10.

We used the same transformations for all libraries to make the benchmarks comparable. The only exception was FFCV, which has its own implementation of the different transformations. For image classification the transformation stack consisted of the following: Random Horizontal Flip, Normalization, Cutout, Transform into Tensor.

For object detection, we relied mostly on Albumentations' \cite{buslaev2020albumentations} implementation of transformations. The stack looked as follows: Random Sized Crop, Random Horizontal Flip, Normalization, Transform into Tensor. These transformations apply to both, images and bounding boxes.

\subsection{Experiment Variants}

When possible, we hosted the dataset locally and in an S3-equivalent bucket. This enabled us to assess the slowdown resulting from training from a stream of data over the network. We will provide a detailed description of the training setting in Section \ref{sec:numerical}.

Given that the most intensive training jobs involve using more than one GPU, whenever possible we also ran the same experiments in an environment with multiple GPU units. Because at the time of running the experiments not all libraries had the full support of PyTorch Lightning, we decided to implement the multi-GPU using  the Distributed Data Parallel (DDP) \cite{li2020pytorch} library from PyTorch.

For some machine learning projects, we might need access only to a subset of a larger dataset. For those cases, having the ability to quickly filter the required data points without having to iterate over the whole dataset can drastically reduce the total training time. Some libraries allow filtering based on certain features, such as the class (for image classification tasks). We explored the gain (or loss) in speed for using the filtering method provided by the library (in case it offered one) versus not filtering at all. Whenever the library did not offer a filtering method, we implemented them naively, i.e., scanning the whole dataset and keeping only those elements that match the specified condition.
Fast filtering is not necessarily trivial to implement as it requires an index-like additional structure to be maintained to avoid iterating over all samples.

Finally, Table \ref{tab:capabilities} specifies the compatibility of the different libraries with the different experiments and datasets we explored in this paper.

\begin{table*}[h]
    \centering
    \begin{tabular}{c|c|c|c|c|c|c|c|c}
         &   & PyTorch & FFCV & Hub & Deep Lake & Torchdata & Webdataset & Squirrel \\
         \hline
         \hline
         CIFAR-10 & default    & Y & Y & Y & Y & Y & Y & Y \\
         CIFAR-10 & remote     & N & N & Y & Y & N & Y & X \\
         CIFAR-10 & filtering  & W & X & Y & Y & Y & Y & Y \\
         CIFAR-10 & multi-GPU  & Y & Y & N & Y & X & X & Y \\
         \hline
         RANDOM & default      & Y & Y & Y & Y & Y & Y & Y \\
         RANDOM & remote       & N & N & Y & Y & N & Y & X \\
         RANDOM & filtering    & W & X & Y & Y & Y & X & X \\
         RANDOM & multi-GPU    & Y & Y & N & Y & X & X & Y \\
         \hline
         CoCo & default        & Y & N & Y & Y & Y & Y & Y \\
         CoCo & remote         & N & N & Y & Y & N & Y & X \\
         CoCo & filtering      & W & X & X & Y & X & X & X \\
         CoCo & multi-GPU      & Y & Y & N & Y & X & X & Y \\
         \hline
    \end{tabular}
    \caption{Comparison of the different libraries and their support for the tested functionalities. (Y)es, supported and implemented. (N)ot supported. (X) not implemented. (W)orkaround found, not supported by default.}
    \label{tab:capabilities}
\end{table*}

\subsection{Metrics}

Our main priority when building the experiments was to find an objective metric that would allow us to compare all the different libraries in a way that was sound.

The ideal metric would have been the total running time of the training job since this is what we have to wait and pay for. Unfortunately, that would have greatly limited the number of experiments we could ran. After careful consideration, we opted for the number of processed data points (images) per second, a result backed by our numerical experiments. We consider two variants of this metric: one in which we use the ML model to train and we perform backpropagation and one in which we do not use the ML model and only iterate over the samples, copying them to GPU. The difference between the two metrics can be appreciated from the pseudo-code of the training loop in Algorithm \ref{alg:training_loop}, where $m$ denotes the speed variable. We also collected the total running time\footnote{This is the time from the start of the simulation until the cutoff, which in practice was often only 10 batches.} and the time it took for the dataloaders to be initialized. The latter was motivated by the fact that some of the libraries might perform expensive computations upfront to increase their speed while training.
We also ended up performing a warm-up for calculating the speed. This is discussed further in Subsection \ref{sec:closer-look}.

\begin{algorithm}[h]
   \caption{The training loop used for computing the speed of the different frameworks.}
   \label{alg:training_loop}
\begin{algorithmic}
   \STATE $t_0 \leftarrow now()$
   \STATE $N \leftarrow 0$ \COMMENT{Total number of data points processed.}
   \STATE $C \leftarrow $ cutoff \COMMENT{When to stop running in seconds.}
   \STATE $m \leftarrow 0$  \COMMENT{Desired metric: samples per second.}
   \FOR{$mode$ {\bfseries in} $(train, val, test)$}
   \STATE $t_0^{d_{mode}} \leftarrow now()$
   \STATE init-dataloader(mode)
   \STATE $t_f^{d_{mode}} \leftarrow now() - t_0^{d_{mode}}$
   \ENDFOR
   \FOR{$e=1,\dots,E$}
   \STATE \COMMENT{Iterate over epochs}
   \STATE $t_0^e \leftarrow 0$
   \FOR{$b$ {\bfseries in} $dataloader$}
   \STATE \COMMENT{Iterate over batches}
   \STATE $X, y \leftarrow b$
   \STATE $t_e \leftarrow \ now() - t_0^e$
   \IF{$C > 0$ and $t_e \geq C$}
   \STATE $m \leftarrow N / t_e$
   \STATE return m
   \ENDIF
   \STATE $N \leftarrow N + X.shape[0]$
   \STATE $data\_to\_gpu(X, y)$
   \IF{$run\_model$}
   \STATE $y_p \leftarrow model(X)$
   \STATE $loss \leftarrow \mathcal{L}(y, y_p)$
   \STATE $loss.backward()$
   \ENDIF
   \ENDFOR
   \STATE $t_f^e \leftarrow now() - t_0^e$
   \ENDFOR
   \STATE $t_f \leftarrow now() - t_0$
\end{algorithmic}
\end{algorithm}

\subsection{Correlation between speed and running time}

\begin{figure}
    \centering
    \includegraphics[width=0.4\textwidth]{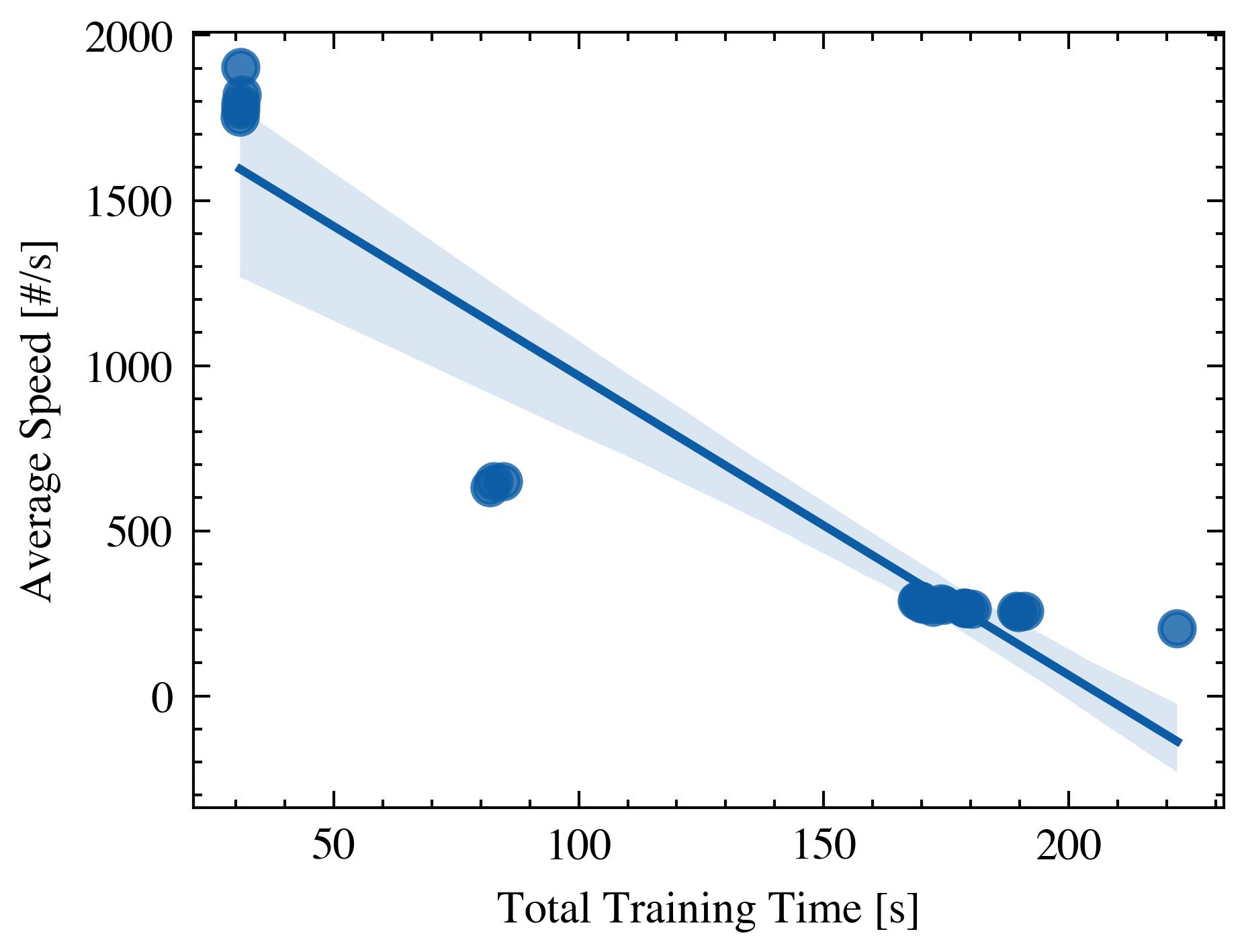}
    \caption{Correlation between Average Speed and Total Training Time,}
    \label{fig:corr_speed_time}
\end{figure}

To justify the usage of speed instead of total running time, our work includes the study of the relationship between the processing speed of each library and the corresponding time elapsed in training.
For the RANDOM dataset, we ran one epoch of each library with a fixed set of parameters twice. From this, we obtained the total training time as well as the speed. Instead of calculating the speed overall for all batches, we used the information available only for the first 10, since that is the only information available in our larger experiments. We saw no difference when limiting the number of batches for calculating the speed, i.e., using 10 instead of all of them.
Supporting our intuition, these values show a strong negative linear correlation between those two variables (Pearson Correlation Coefficient of $-0.93$, p-value=$1.3e^{-11}$) (see Figure \ref{fig:corr_speed_time}).

\subsection{A closer look into running times}
\label{sec:closer-look}

\begin{figure}[h]
    \centering
    \includegraphics[width=0.5\textwidth]{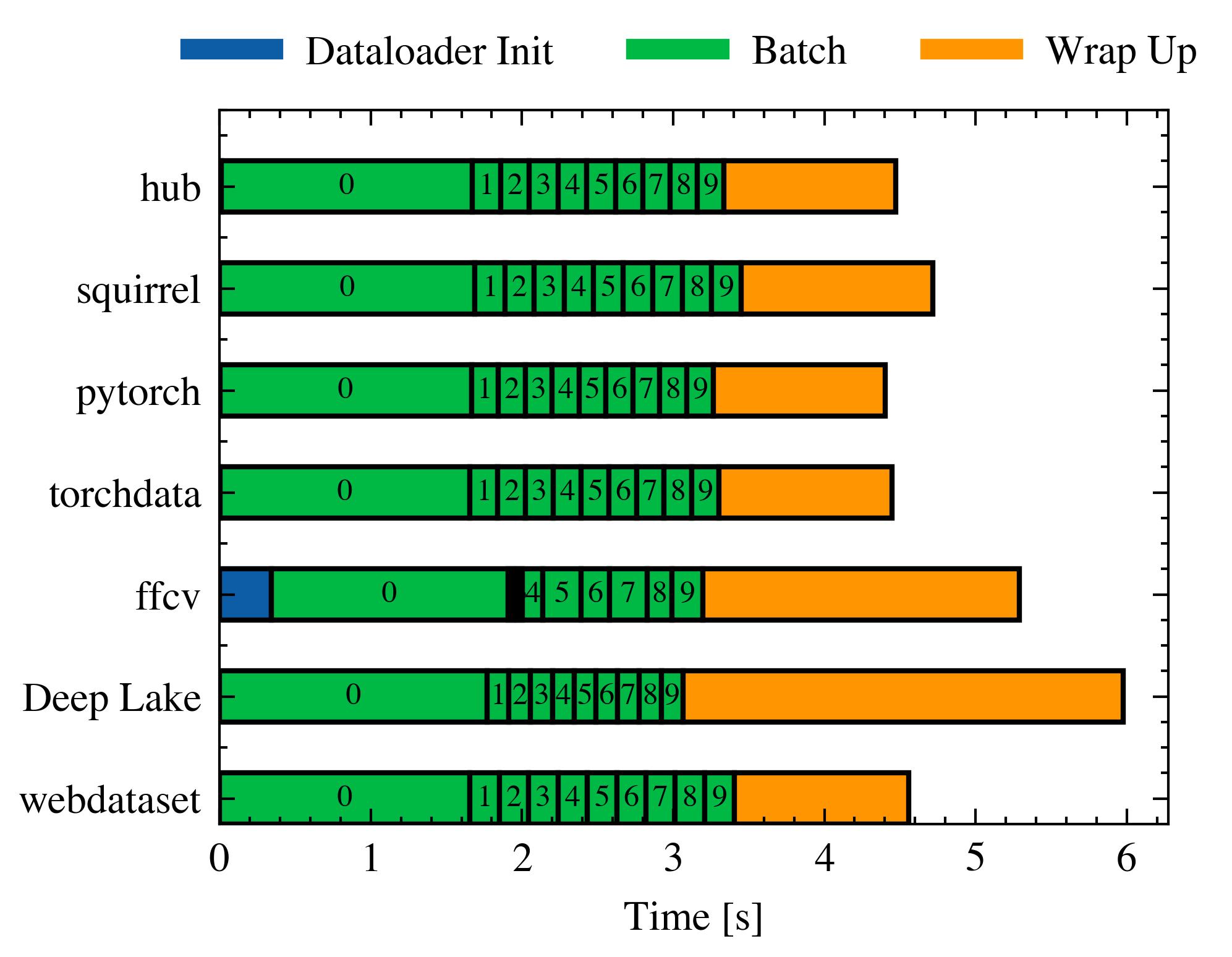}
    \caption{Inspecting the time taken by each library for a single simulation.}
    \label{fig:closer_look}
\end{figure}

To increase our understanding of the inner mechanisms in each library, we decided to inspect for a single run how long it took to execute each batch as well as to initialize the dataloader.
Figure \ref{fig:closer_look} depicts for a single combination of parameters \footnote{RANDOM dataset, single GPU, 0 workers, batch size 64}, the time taken by each library in the steps described by Algorithm \ref{alg:training_loop}. All these experiments involved a cutoff after 10 batches.

Interestingly, the first batch takes a considerable time longer than the others. This can be explained as follows: since most dataloaders rely on lazy loading data at this point, future calls will benefit from pre-fetching, data already in memory, and parallelization (doing things while the GPU is busy doing computations).

The size of the bands 1 to 9 provides the best indication of how well each library scales since the time taken on a large dataset grows linearly with that width. We can observe that most libraries have a uniform width, with Deep Lake being the shortest (the fastest). On the other hand, the only library that shows non-homogeneous widths is FFCV, where the bands 1 to 3 are so thin that they cannot be seen in the image. 

The wrap-up period takes about the same time for all libraries except for FFCV and Deep Lake, which take considerably longer. The time spent wrapping up depends mostly on the model and is not necessarily indicative of how well each library scales. 

Based on this figure, we decided to perform a warm-up when computing the speed. This translates into ignoring the time taken by the first batch in all speed calculations.

\section{Numerical Results}\label{sec:numerical}

For the first set of experiments, we evaluated the performance of all libraries when changing the number of workers (see \ref{sec:dataloaders}) as well as the batch size. These experiments were run in a local server with the following specifications: Intel(R) Core(TM) i9-10900K, 2 x NVIDIA GeForce RTX 3090, and an HDD with 10TB of storage (6GB/s) \footnote{https://www.seagate.com/www-content/product-content/barracuda-fam/barracuda-new/en-us/docs/100835983b.pdf}. 

We evaluated the three modes: default (single GPU), distributed (two GPUs), and filtering (single GPU) for all possible combinations of 0, 1, and 2 workers. For CIFAR10 and RANDOM, the batch size was either 16, 64, or 128. For CoCo, we used smaller values: 1, 2, and 4. All these experiments used a cutoff of 10 and the variant that runs the model (forward \& backward pass).

\subsection{Speed as a function of batch size and workers}

\begin{figure}
    \centering
    \includegraphics{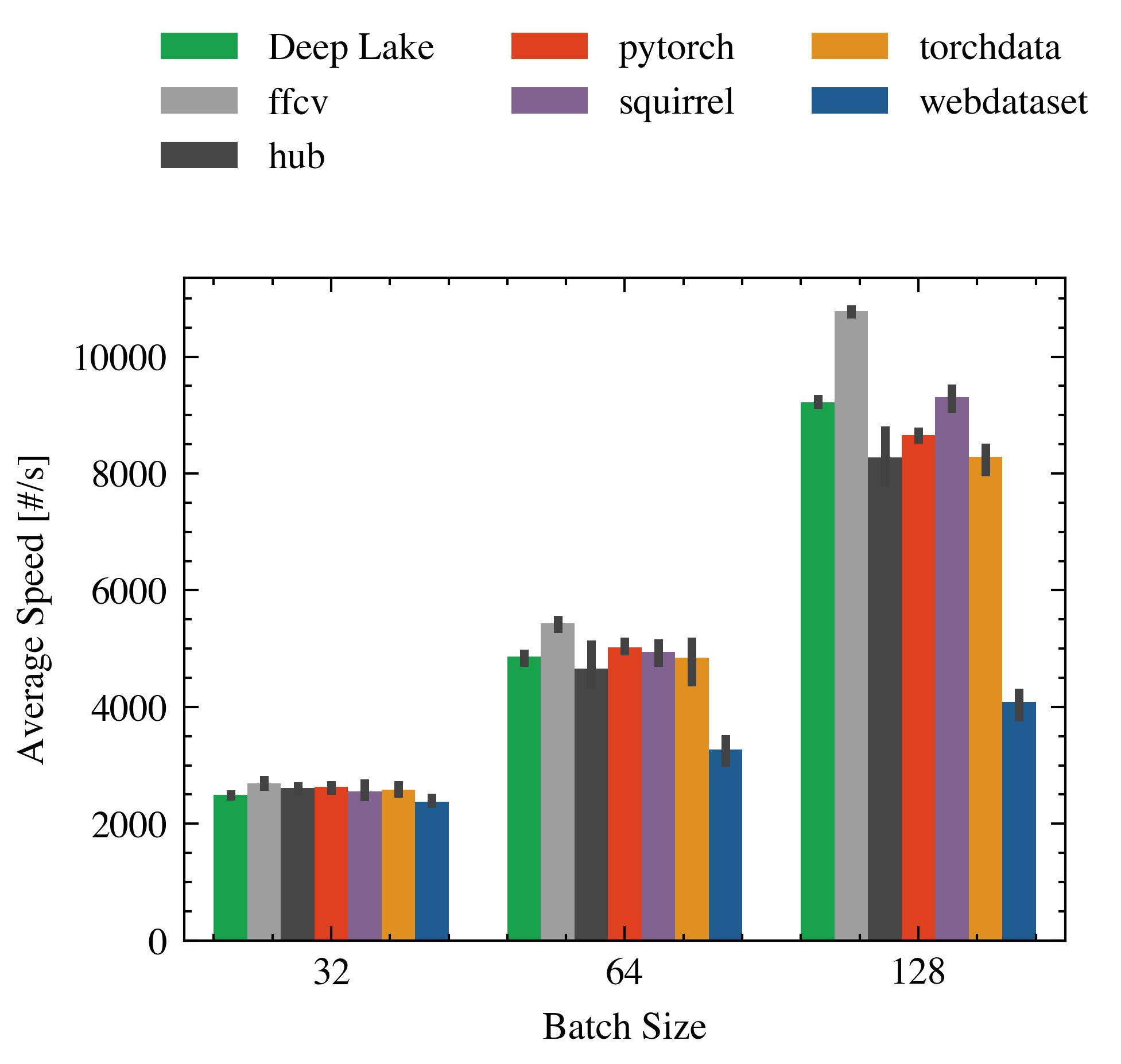}
    \caption{Speed as a function of the batch size for CIFAR10 on a single GPU.}
    \label{fig:speed_batch}
\end{figure}

\begin{figure}
    \centering
    \includegraphics{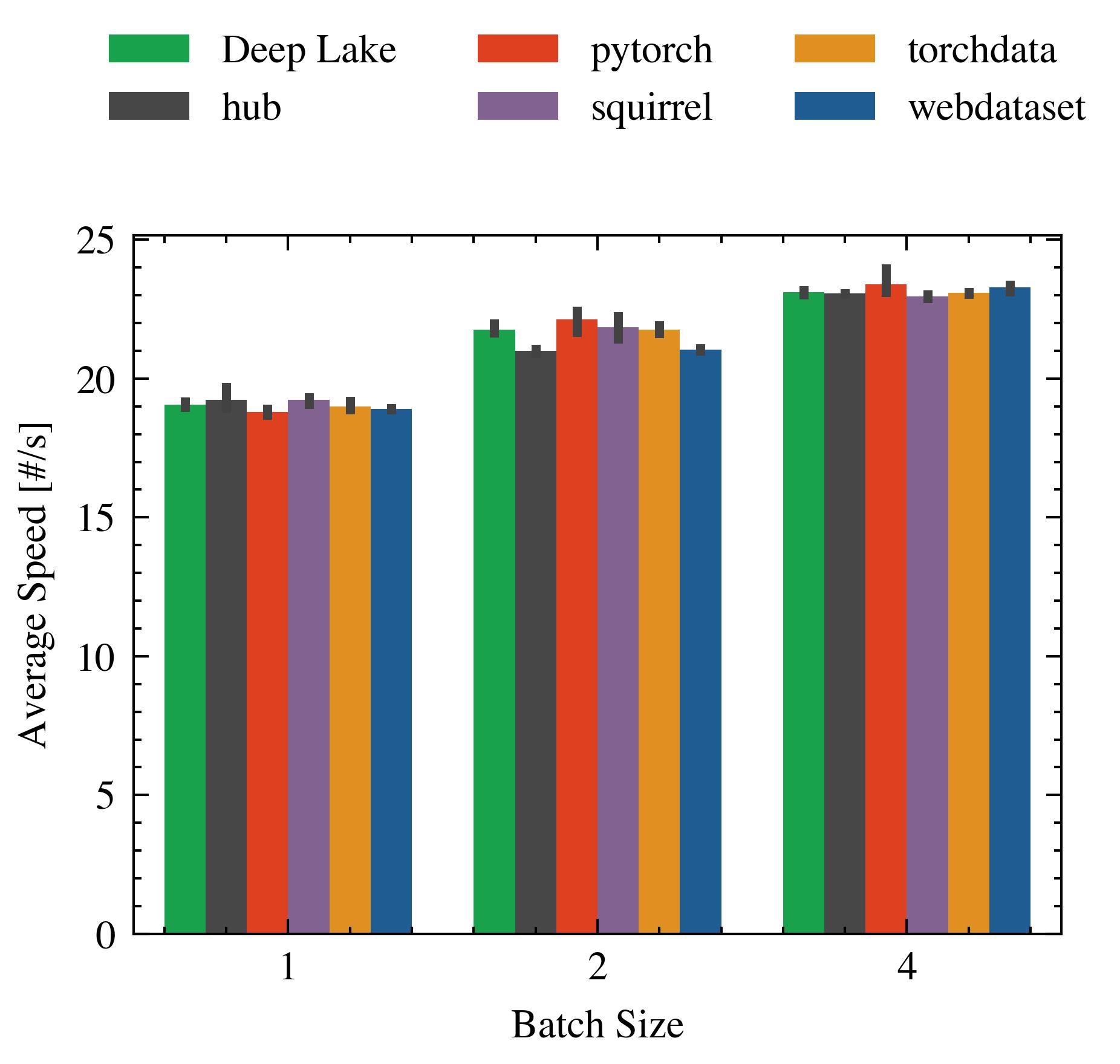}
    \caption{Speed as a function of the batch size for CoCo on a single GPU.}
    \label{fig:speed_batch_coco}
\end{figure}

The first thing we notice while examining the experiments is that the variance between libraries depends on the problem and the dataset used. Figure \ref{fig:speed_batch} shows one such comparison for CIFAR10 on a single GPU, whereas \ref{fig:speed_batch_coco} show the same result for CoCo, also on a single GPU.

This was to be expected given that in the latter, the time taken to compute the forward and backward pass dominates the overall running time, which is not the case for image classification. You may also note the overall difference in speed: from 4000 samples per second to only 20.

\begin{figure}
    \centering
    \includegraphics{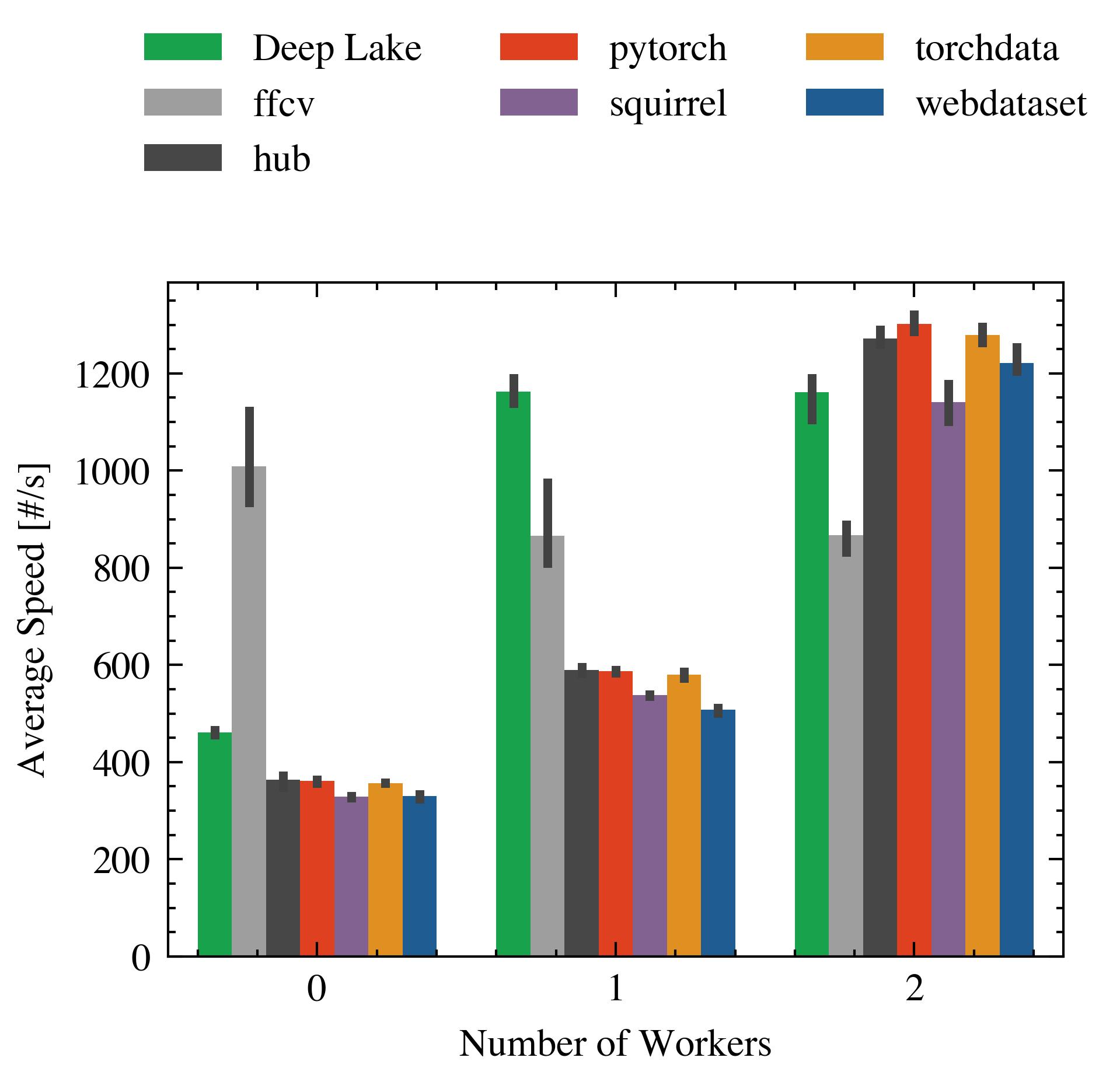}
    \caption{Speed as a function of the number of workers for RANDOM on a single GPU.}
    \label{fig:speed_workers}
\end{figure}

Secondly, we point out that increasing the batch size increases processing speed in almost all cases. However, this is not the case for the number of workers. We can observe in Figure \ref{fig:speed_workers} that FFCV performance degrades as the number of workers increases, while Deep Lake plateaus at 1 worker and not 2. One explanation is that the libraries have their own internal algorithms that decide how to span threads and processes as needed. The point above is relevant for users of these libraries, as experience with one of them might not translate well into another one.

\subsection{Speed Gains when using DDP}

\begin{figure}[h]
    \centering
    \includegraphics{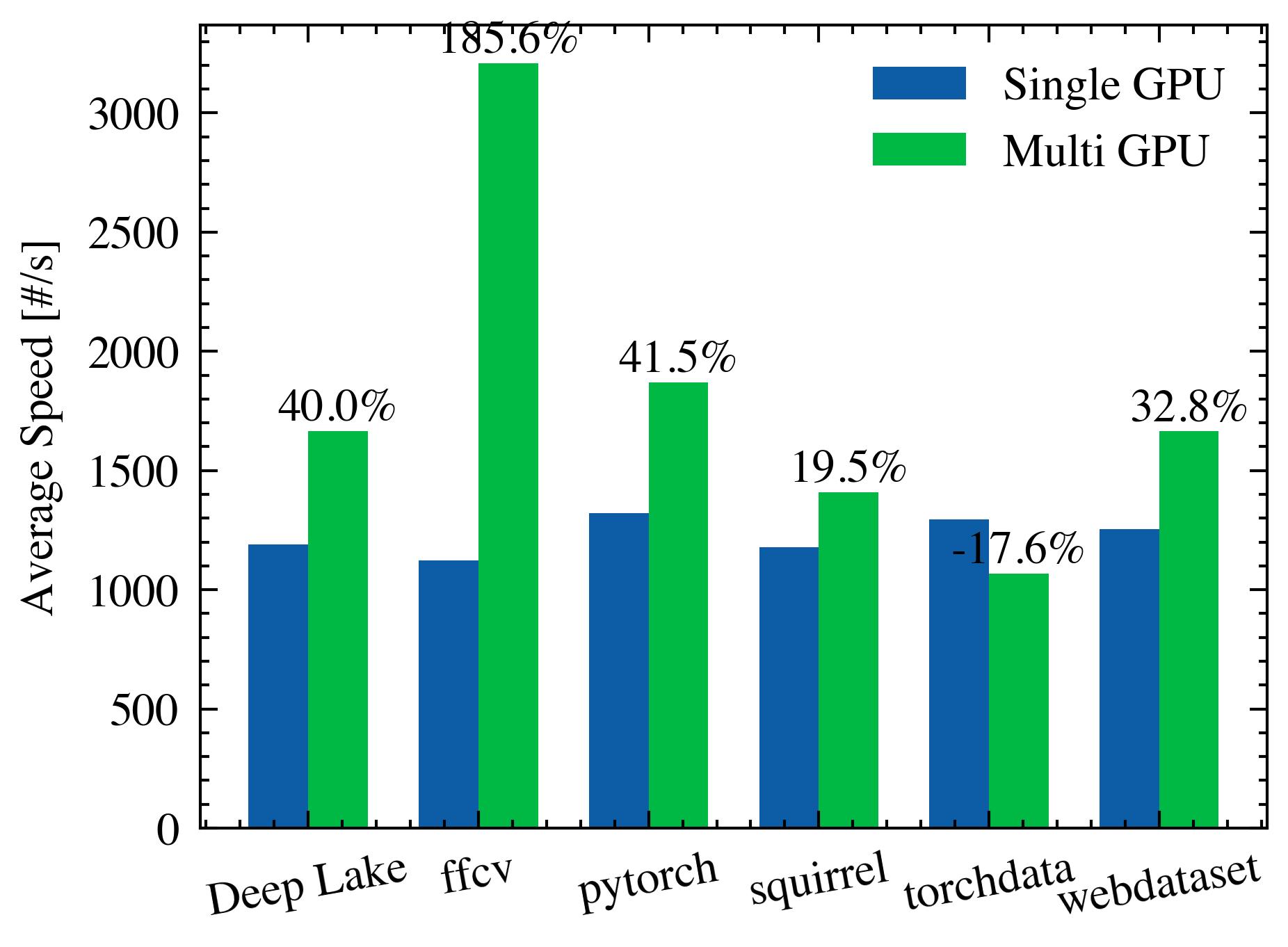}
    \caption{Maximum Speed gains when using 2 GPUs over one for the RANDOM dataset.}
    \label{fig:ddp_random}
\end{figure}

A desirable feature of a dataloader is its ability to scale linearly with the number of GPUs. This is not always possible and depends on the internal loading mechanism of each library. We explore how these libraries performed by comparing the speed increase when using one or two GPUs. Figure \ref{fig:ddp_random} shows the results for the RANDOM dataset. Each bar represents the maximum speed achieved across all batch sizes, number of workers, and repetitions. In a way, this reflects the maximum speed achievable by the library. We observe that libraries speed up about 40\%, less than half of a linear increase on average. Two cases are particularly surprising. On the one hand, Torchdata performs worse with two GPUs than on a single one. On the other hand, FFCV achieved a speed increase of more than theoretically possible. 
There are several artifacts that can be at play here, but most likely, it is due to the limited number of repetitions we were able to run (due to limited resources). Also, this might indicate a misconfiguration in Torchdata: the documentation on running experiments in multi-GPU environments is limited for most libraries. 

\subsection{Comparison between with and without forward and backward pass}

As we discussed when presenting Algorithm \ref{alg:training_loop}, we had to decide whether we would incorporate the forward and backward passes into the speed calculation. There are arguments for both. On the one hand, including the forward and backward passes better reflect the algorithm's actual training time. At the same time, some libraries might preemptively optimize steps normally done during the forward pass, so stopping there would seem as if they take longer than they do.

On the other hand, if the time taken by the forward and backward pass is much larger than the time it takes for only loading the data, including that time in the calculation would inevitably hide the difference between libraries.

\begin{figure}
    \centering
    \includegraphics{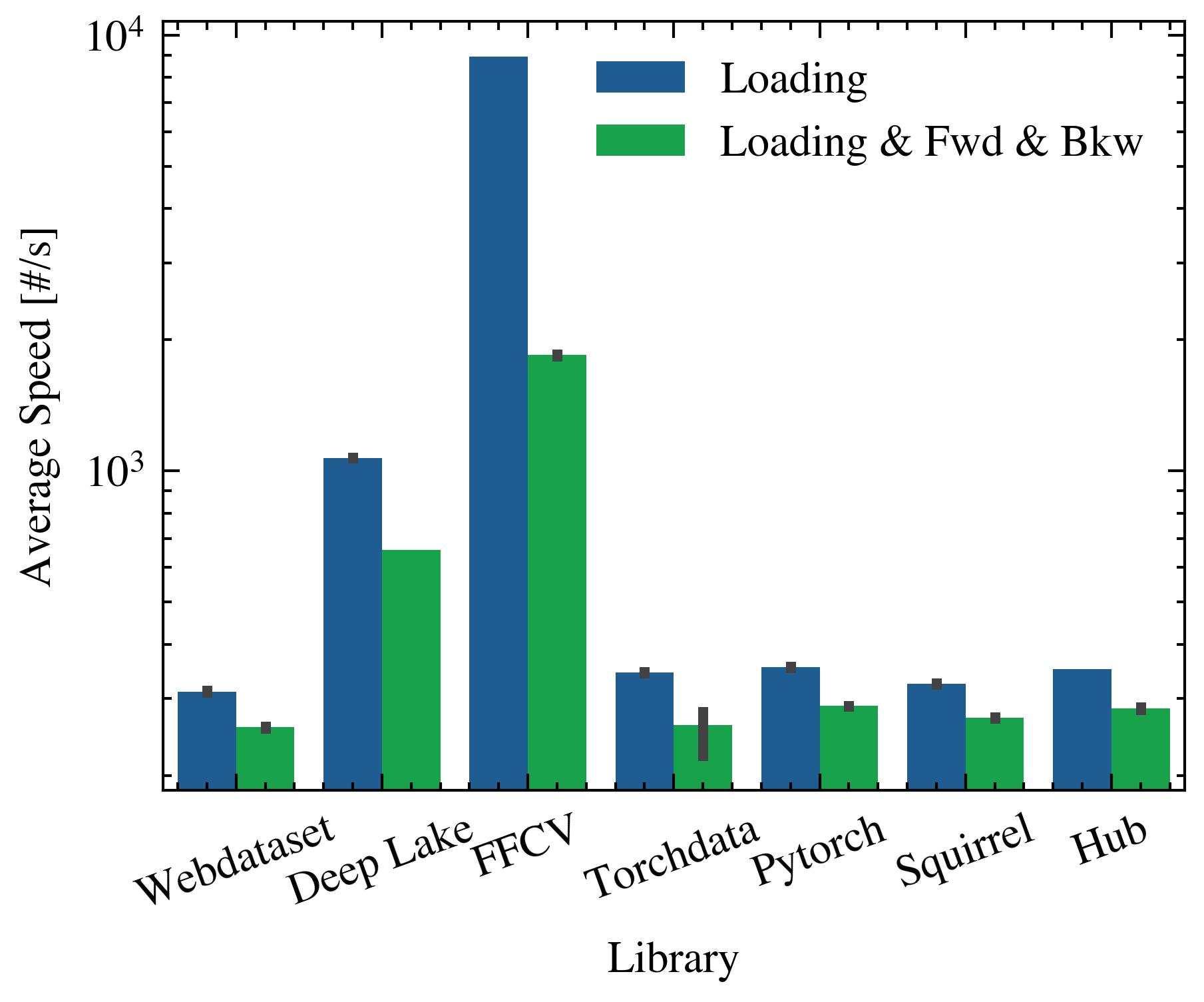}
    \caption{Difference in training time when computing the forward and backward passes and when not. Y-axis is on log scale.}
    \label{fig:run_model}
\end{figure}

To understand if the behavior change was noticeable, we used the RANDOM dataset to compare the difference in average speed when including the two model operations in the calculation and when not. The results are presented in Figure \ref{fig:run_model}. We can observe that most libraries have a slightly increased speed when excluding the model (except for FFCV, whose performance drops in half), and most importantly, the relative performance among libraries remains almost the same.

\subsection{Speed trade-offs when filtering data}

For our filtering experiments, we selected two classes to keep for CIFAR10 and RANDOM: \emph{dog} and \emph{truck}, and \emph{0} and \emph{13}, respectively. For CoCO we selected three classes: \emph{pizza}, \emph{couch}, \emph{cat}.

We observed that most libraries do not have a good filtering mechanism that avoids iterating over the whole dataset. For example, our PyTorch filtering implementation requires building a custom sampler with the indices of the desired images.

\begin{figure}
    \centering
    \includegraphics{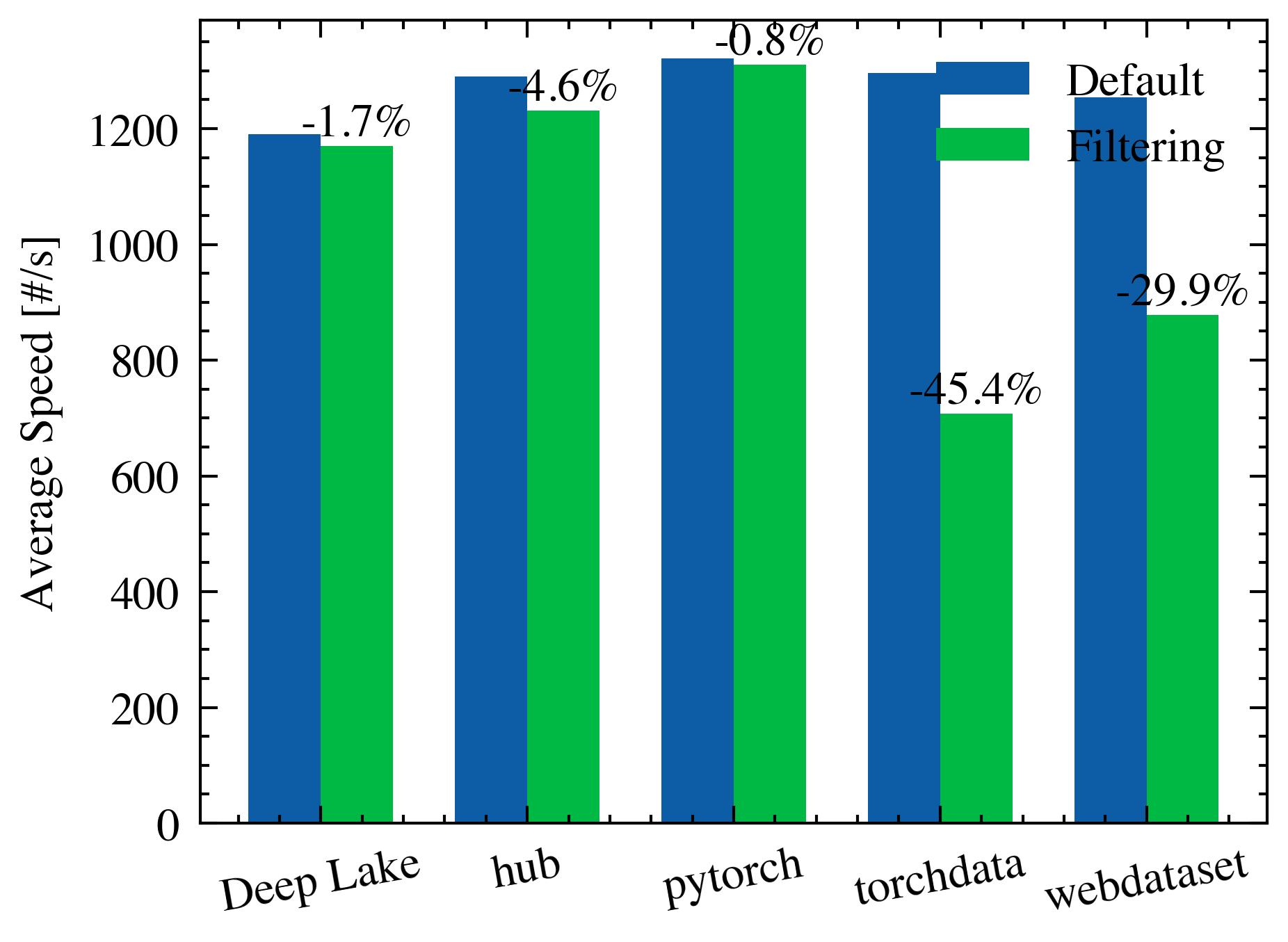}
    \caption{Speed losses when filtering the RANDOM dataset.}
    \label{fig:filtering_random}
\end{figure}

This is quite fast for a small dataset but becomes unfeasible for large datasets: filtering CoCo using PyTorch was prohibitively expensive. In general, performance was quite similar when filtering and when not\footnote{in speed. Building the sampler for PyTorch is done upfront, considerably affecting the total running time.}. Similarly to Figure \ref{fig:ddp_random}, in Figure \ref{fig:filtering_random}, we can see the slowdown as a result of filtering: even though it was considerable for Torchdata and Webdataset, we saw a reversal of the results when working with the CoCo Dataset.

\subsection{Training while streaming over the network}

Ideally, we could decouple the dataset storage from the machine learning training process and simply connect the database storing our data to the ML framework of choice, regardless of where the two are located. That involves sending the training data over a network and losing considerable speed. With the high costs involved in renting GPU accelerated hardware on the cloud, it might seem that the convenience is not worth it. But is it not?

Some of the libraries considered in this paper allow specifying a dataset accessible via the internet: Webdataset, Hub, and Deep Lake are particularly good at this\footnote{Squirrel has this capability, but we did not manage to specify a MinIO address, so we excluded it from the comparison. We had a similar issue with Torchdata.}. The question then becomes: how big is the tradeoff between ease of usage and running time?

We set up the following experiment to offer some insight into this question. We ran two full epochs of the RANDOM dataset for the three libraries: Hub, Deep Lake, and Webdataset, while changing the origin of the data. Three locations were considered: a local copy in the machine running the experiments' hard drive, a copy in an S3 bucket (in the closest region to our machine), and a copy stored in MinIO (an open source equivalent of S3 that can be hosted locally) running in a similar machine within the same local network (both machines were connected via WiFi). The experiments' computer had an Intel(R) Core(TM) i7-7700 CPU @ 3.60GHz, 16 GB of RAM, NVIDIA GeForce RTX 2070 Rev, and a Samsung SSD 850 hard drive with 256 GB of storage. Regarding latency, the Round Trip Time from the workstation running the experiments to the MinIO server (located in the same room and in the same local WiFi) took {$59.2 \pm 58.5ms$} (min. {$8.8ms$}), while to the S3-bucket in AWS servers took {$17.3 \pm 1.3ms$} (min. {$14.8ms$}).


\begin{figure}
    \centering
    \includegraphics{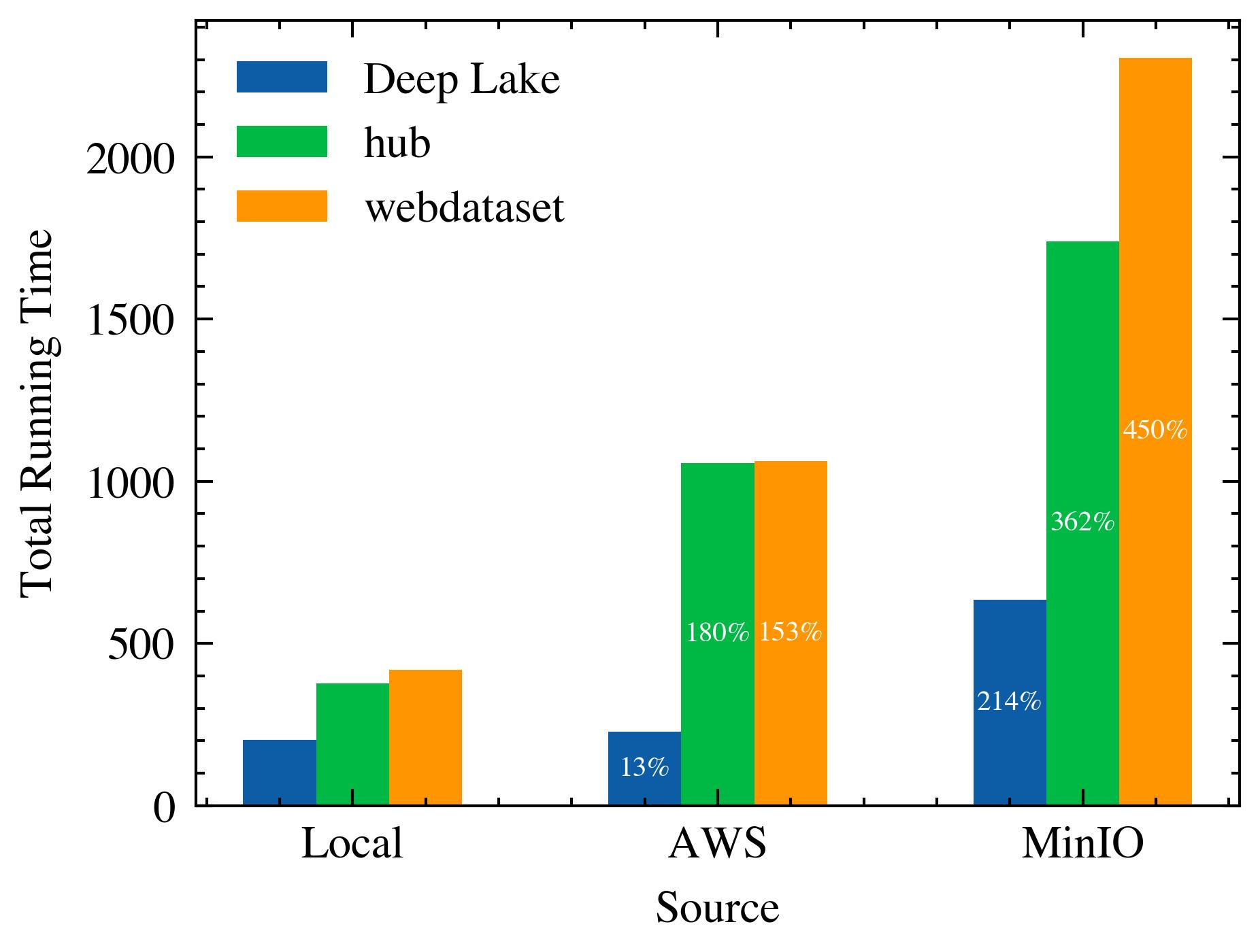}
    \caption{Comparing the performance of Hub, Deep Lake, and Webdataset when loading data from different locations: Local, AWS, and MinIO.}
    \label{fig:location_location}
\end{figure}

Figure \ref{fig:location_location} depicts the total running times for the nine experiments, and the white percentages denote the slowdown (increase in running time) compared to the local case. We can observe that even though for Hub and Webdataset there is a significant increase when moving to AWS, Deep Lake managed to maintain almost the same speed with an increase of only 13\%. Another helpful insight could be extracted from the result: the MinIO setting shows a slowdown almost twice as bad as the AWS setting, as seen in Figure \ref{fig:location_location}. This output could be explained primarily by the difference in average Round Trip Times shown above, highlighting that local networks (e.g., internal company networks\footnote{In this case a university network}) might not be the most efficient way to host datasets due to their complex configurations and restrictions. Furthermore, this result also indicates that the storage serving the datasets over the network plays a crucial role in enabling remote training and might trigger questions on best practices to serve datasets. Namely, data from S3 is served in parallel from different servers with load balancing while we had access to a single MinIO instance.

The plots for all additional experiments can be found in Appendix \ref{ap:extra-experiments}.

\section{Discussion}\label{sec:discussion}

In this work, we used time as the main tool to compare the performance among different libraries. There are several things to say about this.
Firstly, we noticed that running times are quite variable and depend on background processes that are hard to control. At the same time, access to multi-GPU resources is expensive, which limits the number of experiments that can be run. Ideally, we would have run more than three repetitions of each experiment with more parameters (more workers, more batch sizes), but we did not have the resources for it. Since we are making all our open-source code, we invite readers to run the benchmarks on their own hardware and report the results. At the same time, libraries are updated fairly often, and a change in version can dramatically increase or decrease its performance.

In light of the above points, we encourage the reader to internalize the qualitative aspects of this paper but beware that the numbers obtained here are prone to change.

Secondly, an aspect that is harder to compare is the ease of use of the libraries considered in this project. Most of the libraries included in this benchmark do not have comprehensive documentation and rely primarily on concrete examples. Consequently, implementation in these libraries is not trivial and prone to inefficiencies. One of the advantages of making our code open source is that we allow any developer to identify and improve on our code. This is particularly relevant as we expect that the benchmarks created in this project could be used as boilerplate code for the community.

We note that there does not seem to be a better library than all others. Instead, each one has its own strengths. Consider the example of FFCV: it seems to be the fastest in our experiments, but the lack of support for label transformations prevents it from being adopted in projects that require such features.  

We hope to analyze the interplay between filtering and training across multiple GPUs in future work. At the same time, it would be interesting to explore the scaling capabilities of these libraries as the number of GPUs increases. Similarly, it would be of great interest to benchmark data loading libraries in terms of performance on the shuffling step in the DL training workflow, as this can have a significant impact on the total training time, and its implementation is a non-trivial problem, where there are several kinds of approaches.

The research on libraries that provide data loading from a remote storage and that they show comparable results with the local storage experiments incentivized us to explore the idea of formulating and designing a caching policy for data streaming over a network. In that setting, reducing the times a data point (e.g., image) needs to be transferred can significantly shorten the overall training time (and possibly costs if network usage is paid). The idea of caching a network dataset while training is not new \cite{mohan2020analyzing}. Still, it is often assumed that the whole dataset can be cached when discussing training and streaming data. Furthermore, it is assumed that all the samples will be used once per epoch \cite{abhishek2020quiver} as is traditionally the case. We are interested in exploring what happens when the cache size is small, and also in removing the requirement of using every datapoint once per epoch.
Such a formulation should borrow from active learning, data summarization, and curriculum learning.


\section{Related Work}\label{sec:relatedwork}

This section describes several efforts in the community to benchmark deep learning libraries, models, and frameworks.

A large body of work exists towards benchmarking deep learning tools and methods. MLPerf \cite{mattson2020mlperf} is arguably the most popular ML benchmarking project for modern ML workloads that targets both training and inference, spanning a variety of AI tasks. The authors use as their objective metric the training time required to reach a given accuracy level. This metric requires increased computational resources and is not well suited for testing dataloader parameters. 
DeepBench \cite{DeepBench} is an open-source project from Baidu Research focused on  kernel-level operations within the deep learning stack; it benchmarks the performance of individual operations (e.g., matrix multiplication) as implemented in libraries and executed directly on the underlying hardware.  Similarly, AI Matrix \cite{zhang2019ai} uses microbenchmarks to cover basic operators, measuring performance for fully connected and other common layers, and matches the characteristics of real workloads by offering synthetic benchmarks. 

\textbf{Comparison of frameworks}:
This section includes efforts toward benchmarking and comparing different deep learning frameworks, such as PyTorch, TensorFlow, etc.

In Deep500 \cite{ben2019modular}, authors provide a modular software framework for measuring DL-training performance; while customizable, it lacks hyperparameter benchmarking and does not provide an easy-to-use way to add and experiment with novel libraries and workflows. AIBench \cite{gao2020aibench}, and DAWNBench \cite{coleman2019analysis} are both end-to-end benchmarks, with the latter being the first multi-entrant benchmark competition to measure the end-to-end performance of deep-learning systems. As with MLPerf, none examine the effect of alternative loading libraries in their workflows. In \cite{wu2019comparative}, the authors present a systematic analysis of the CPU and memory usage patterns for different parallel computing libraries and batch sizes and their impact on accuracy and training efficiency. This analysis is close to our work; however, it does not provide an open-source resource to interact with and benchmark new libraries.

In \cite{shi2016benchmarking}, the authors compare deep learning frameworks based on the performance of different neural networks (e.g., Fully Connected, Convolutional, and Recurrent Neural Networks). dPRO \cite{hu2022dpro} focuses on distributed (multi-GPU) training benchmarks by utilizing a profiler that collects runtime traces of distributed DNN training across multiple frameworks. DLBench \cite{DLBench} is a benchmark framework for measuring different deep learning tools, such as Caffe, Tensorflow and MXNet. In \cite{liu2018benchmarking} authors study the impact of default configurations by each framework on the model performance (time and accuracy), demonstrating the complex interactions of the DNN parameters and hyperparameters with dataset-specific characteristics. Yet, the experiments include only the default configurations of each framework and lack any analysis of non-default settings. In \cite{wu2018experimental}, the authors test default configurations of frameworks and attempt to find the optimal ones for each dataset; they also examine the data loading process but do not evaluate third-party libraries. 
All the previously published works in this paragraph, while they bear numerous similarities with our work, they have one significant distinction with it; they do not conduct any analysis or benchmarking on PyTorch or the ecosystem of libraries for data loading described in this paper, which, as stated in the introduction, is currently one of the most popular deep learning frameworks that are widely utilized both in industry and academia.

\textbf{Comparison of different DNN architectures and hardware}: ParaDNN \cite{wang2020systematic} generates parameterized end-to-end models to run on target platforms, such as varying the batch size to challenge the bounds of the underlying hardware, but focuses on the comparison of specialized platforms (TPU v2/v3) and device architectures (TPU, GPU, CPU). Relevant to ParaDNN is the work of \cite{bianco2018benchmark}, which provides a comprehensive tool for selecting the appropriate architecture responding to resource constraints in practical deployments and applications based on analysis of hardware systems with diverse computational resources. However, it concentrates more on the design of deep learning models than the deep learning frameworks these are implemented on. While Fathom \cite{adolf2016fathom} and TBD Suite \cite{zhu2018tbd} both focus on the evaluation of full model architectures across a broad variety of tasks and diverse workloads, they are limited on these and lack benchmarks for state-of-the-art training innovations.


\textbf{Other Devices:} AI Benchmark \cite{ignatov2018ai} is arguably the first mobile-inference benchmark suite. However, its results focus solely on Android smartphones and only measure latency while providing a summary score that explicitly fails to specify quality targets. \cite{hadidi2019characterizing} investigates the in-the-edge inference of DNNs from execution time, energy consumption, and temperature perspectives. \cite{tao2018benchip} covers configurations with diverse hardware behaviors, such as branch prediction rates and data reuse distances, and evaluates the accuracy, performance, and energy of intelligence processors and hardware platforms. Both of these works are fixated on a different range of devices, such as edge devices and intelligence processors, which is out of the scope of this work. 


\section{Conclusions}\label{sec:conclusions}

In this paper, we explored the current landscape of Pytorch libraries that allow machine learning practitioners to load their datasets into their models. These libraries offer a wide array of features from increased speed, creating views of only a subset of the data, and loading data from remote storage. We believe that remote loading holds the most promise for all these features since it enables the de-coupling of data storage and model training. Even though loading speed over the public internet is naturally slower than from a local disk, some libraries, such as Deep Lake, showed remarkable results (only a 13\% increase in time).
For the most part, we did not find a considerable difference in performance across libraries except for FFCV for multi-GPUs and Deep Lake for networked loading, which performed remarkably well. However, we did notice that the documentation for most of these libraries is not readily available or comprehensive, which might result in misconfigured setups. Since good practices are hard to find, a programmer might use what works well in a different dataloader, which need not work in the new library. At this point, the performance gains do not seem large enough to justify the migration of existing code bases for small to medium jobs. For larger jobs, there could be significant cost reductions for switching to one of the faster libraries.
Finally, we believe that an innovative caching system designed for machine learning applications could be the final piece in realizing the vision of a truly decoupled dataset model system. Any such approach would have to build existing knowledge on dataset summarization and active learning.

\section*{Acknowledgements}

The authors would like to thank the Activeloop team for their support and insights during the development of this project. The authors would also like to thank both Tryolabs and Activeloop for their resources for running some of the experiments.




\bibliography{paper}

\begin{thebibliography}{35}
\providecommand{\natexlab}[1]{#1}
\providecommand{\url}[1]{\texttt{#1}}
\expandafter\ifx\csname urlstyle\endcsname\relax
  \providecommand{\doi}[1]{doi: #1}\else
  \providecommand{\doi}{doi: \begingroup \urlstyle{rm}\Url}\fi

\bibitem[Abadi et~al.(2015)Abadi, Agarwal, Barham, Brevdo, Chen, Citro,
  Corrado, Davis, Dean, Devin, Ghemawat, Goodfellow, Harp, Irving, Isard, Jia,
  Jozefowicz, Kaiser, Kudlur, Levenberg, Man\'{e}, Monga, Moore, Murray, Olah,
  Schuster, Shlens, Steiner, Sutskever, Talwar, Tucker, Vanhoucke, Vasudevan,
  Vi\'{e}gas, Vinyals, Warden, Wattenberg, Wicke, Yu, and
  Zheng]{tensorflow2015-whitepaper}
Abadi, M., Agarwal, A., Barham, P., Brevdo, E., Chen, Z., Citro, C., Corrado,
  G.~S., Davis, A., Dean, J., Devin, M., Ghemawat, S., Goodfellow, I., Harp,
  A., Irving, G., Isard, M., Jia, Y., Jozefowicz, R., Kaiser, L., Kudlur, M.,
  Levenberg, J., Man\'{e}, D., Monga, R., Moore, S., Murray, D., Olah, C.,
  Schuster, M., Shlens, J., Steiner, B., Sutskever, I., Talwar, K., Tucker, P.,
  Vanhoucke, V., Vasudevan, V., Vi\'{e}gas, F., Vinyals, O., Warden, P.,
  Wattenberg, M., Wicke, M., Yu, Y., and Zheng, X.
\newblock {TensorFlow}: Large-scale machine learning on heterogeneous systems,
  2015.
\newblock URL \url{https://www.tensorflow.org/}.
\newblock Software available from tensorflow.org.

\bibitem[Adolf et~al.(2016)Adolf, Rama, Reagen, Wei, and
  Brooks]{adolf2016fathom}
Adolf, R., Rama, S., Reagen, B., Wei, G.-Y., and Brooks, D.
\newblock Fathom: Reference workloads for modern deep learning methods.
\newblock In \emph{2016 IEEE International Symposium on Workload
  Characterization (IISWC)}, pp.\  1--10. IEEE, 2016.

\bibitem[Baidu-Research(2020)]{DeepBench}
Baidu-Research.
\newblock Deep{B}ench, 2020.
\newblock URL \url{https://github.com/baidu-research/DeepBench}.

\bibitem[Ben-Nun et~al.(2019)Ben-Nun, Besta, Huber, Ziogas, Peter, and
  Hoefler]{ben2019modular}
Ben-Nun, T., Besta, M., Huber, S., Ziogas, A.~N., Peter, D., and Hoefler, T.
\newblock A modular benchmarking infrastructure for high-performance and
  reproducible deep learning.
\newblock In \emph{2019 IEEE International Parallel and Distributed Processing
  Symposium (IPDPS)}, pp.\  66--77. IEEE, 2019.

\bibitem[Bianco et~al.(2018)Bianco, Cadene, Celona, and
  Napoletano]{bianco2018benchmark}
Bianco, S., Cadene, R., Celona, L., and Napoletano, P.
\newblock Benchmark analysis of representative deep neural network
  architectures.
\newblock \emph{IEEE access}, 6:\penalty0 64270--64277, 2018.

\bibitem[Buslaev et~al.(2020)Buslaev, Iglovikov, Khvedchenya, Parinov,
  Druzhinin, and Kalinin]{buslaev2020albumentations}
Buslaev, A., Iglovikov, V.~I., Khvedchenya, E., Parinov, A., Druzhinin, M., and
  Kalinin, A.~A.
\newblock Albumentations: fast and flexible image augmentations.
\newblock \emph{Information}, 11\penalty0 (2):\penalty0 125, 2020.

\bibitem[Coleman et~al.(2019)Coleman, Kang, Narayanan, Nardi, Zhao, Zhang,
  Bailis, Olukotun, R{\'e}, and Zaharia]{coleman2019analysis}
Coleman, C., Kang, D., Narayanan, D., Nardi, L., Zhao, T., Zhang, J., Bailis,
  P., Olukotun, K., R{\'e}, C., and Zaharia, M.
\newblock Analysis of dawnbench, a time-to-accuracy machine learning
  performance benchmark.
\newblock \emph{ACM SIGOPS Operating Systems Review}, 53\penalty0 (1):\penalty0
  14--25, 2019.

\bibitem[Gao et~al.(2020)Gao, Tang, Zhan, Lan, Luo, Wang, Dai, Cao, Xiong,
  Jiang, et~al.]{gao2020aibench}
Gao, W., Tang, F., Zhan, J., Lan, C., Luo, C., Wang, L., Dai, J., Cao, Z.,
  Xiong, X., Jiang, Z., et~al.
\newblock Aibench: An agile domain-specific benchmarking methodology and an ai
  benchmark suite.
\newblock \emph{arXiv preprint arXiv:2002.07162}, 2020.

\bibitem[Hadidi et~al.(2019)Hadidi, Cao, Xie, Asgari, Krishna, and
  Kim]{hadidi2019characterizing}
Hadidi, R., Cao, J., Xie, Y., Asgari, B., Krishna, T., and Kim, H.
\newblock Characterizing the deployment of deep neural networks on commercial
  edge devices.
\newblock In \emph{2019 IEEE International Symposium on Workload
  Characterization (IISWC)}, pp.\  35--48. IEEE, 2019.

\bibitem[Hambardzumyan et~al.(2022)Hambardzumyan, Tuli, Ghukasyan, Rahman,
  Topchyan, Isayan, Harutyunyan, Hakobyan, Stranic, and
  Buniatyan]{deeplake2022}
Hambardzumyan, S., Tuli, A., Ghukasyan, L., Rahman, F., Topchyan, H., Isayan,
  D., Harutyunyan, M., Hakobyan, T., Stranic, I., and Buniatyan, D.
\newblock Deep lake: a lakehouse for deep learning, 2022.
\newblock URL \url{https://arxiv.org/abs/2209.10785}.

\bibitem[Heterogeneous Computing Lab~at HKBU(2017)]{DLBench}
Heterogeneous Computing Lab~at HKBU, D.
\newblock {DLB}ench, 2017.
\newblock URL \url{https://github.com/hclhkbu/dlbench}.

\bibitem[Hinton et~al.(2012)Hinton, Srivastava, and Swersky]{hinton2012neural}
Hinton, G., Srivastava, N., and Swersky, K.
\newblock Neural networks for machine learning lecture 6a overview of
  mini-batch gradient descent.
\newblock \emph{Cited on}, 14\penalty0 (8):\penalty0 2, 2012.

\bibitem[Hu et~al.(2022)Hu, Jiang, Zhong, Peng, Wu, Zhu, Lin, and
  Guo]{hu2022dpro}
Hu, H., Jiang, C., Zhong, Y., Peng, Y., Wu, C., Zhu, Y., Lin, H., and Guo, C.
\newblock dpro: A generic performance diagnosis and optimization toolkit for
  expediting distributed dnn training.
\newblock \emph{Proceedings of Machine Learning and Systems}, 4:\penalty0
  623--637, 2022.

\bibitem[Ignatov et~al.(2018)Ignatov, Timofte, Chou, Wang, Wu, Hartley, and
  Van~Gool]{ignatov2018ai}
Ignatov, A., Timofte, R., Chou, W., Wang, K., Wu, M., Hartley, T., and
  Van~Gool, L.
\newblock Ai benchmark: Running deep neural networks on android smartphones.
\newblock In \emph{Proceedings of the European Conference on Computer Vision
  (ECCV) Workshops}, pp.\  0--0, 2018.

\bibitem[Krizhevsky et~al.(2009)Krizhevsky, Hinton,
  et~al.]{krizhevsky2009learning}
Krizhevsky, A., Hinton, G., et~al.
\newblock Learning multiple layers of features from tiny images.
\newblock 2009.

\bibitem[Kumar \& Sivathanu(2020)Kumar and Sivathanu]{abhishek2020quiver}
Kumar, A.~V. and Sivathanu, M.
\newblock Quiver: An informed storage cache for deep learning.
\newblock In \emph{18th USENIX Conference on File and Storage Technologies
  (FAST 20)}, pp.\  283--296, Santa Clara, CA, February 2020. USENIX
  Association.
\newblock ISBN 978-1-939133-12-0.
\newblock URL
  \url{https://www.usenix.org/conference/fast20/presentation/kumar}.

\bibitem[Leclerc et~al.(2022)Leclerc, Ilyas, Engstrom, Park, Salman, and
  Madry]{leclerc2022ffcv}
Leclerc, G., Ilyas, A., Engstrom, L., Park, S.~M., Salman, H., and Madry, A.
\newblock ffcv.
\newblock \url{https://github.com/libffcv/ffcv/}, 2022.
\newblock commit xxxxxxx.

\bibitem[Li et~al.(2020)Li, Zhao, Varma, Salpekar, Noordhuis, Li, Paszke,
  Smith, Vaughan, Damania, et~al.]{li2020pytorch}
Li, S., Zhao, Y., Varma, R., Salpekar, O., Noordhuis, P., Li, T., Paszke, A.,
  Smith, J., Vaughan, B., Damania, P., et~al.
\newblock Pytorch distributed: Experiences on accelerating data parallel
  training.
\newblock \emph{arXiv preprint arXiv:2006.15704}, 2020.

\bibitem[Lin et~al.(2014)Lin, Maire, Belongie, Hays, Perona, Ramanan,
  Doll{\'a}r, and Zitnick]{lin2014microsoft}
Lin, T.-Y., Maire, M., Belongie, S., Hays, J., Perona, P., Ramanan, D.,
  Doll{\'a}r, P., and Zitnick, C.~L.
\newblock Microsoft coco: Common objects in context.
\newblock In \emph{European conference on computer vision}, pp.\  740--755.
  Springer, 2014.

\bibitem[Liu et~al.(2018)Liu, Wu, Wei, Cao, Sahin, and
  Zhang]{liu2018benchmarking}
Liu, L., Wu, Y., Wei, W., Cao, W., Sahin, S., and Zhang, Q.
\newblock Benchmarking deep learning frameworks: Design considerations, metrics
  and beyond.
\newblock In \emph{2018 IEEE 38th International Conference on Distributed
  Computing Systems (ICDCS)}, pp.\  1258--1269. IEEE, 2018.

\bibitem[Mattson et~al.(2020)Mattson, Cheng, Diamos, Coleman, Micikevicius,
  Patterson, Tang, Wei, Bailis, Bittorf, et~al.]{mattson2020mlperf}
Mattson, P., Cheng, C., Diamos, G., Coleman, C., Micikevicius, P., Patterson,
  D., Tang, H., Wei, G.-Y., Bailis, P., Bittorf, V., et~al.
\newblock Mlperf training benchmark.
\newblock \emph{Proceedings of Machine Learning and Systems}, 2:\penalty0
  336--349, 2020.

\bibitem[Mohan et~al.(2020)Mohan, Phanishayee, Raniwala, and
  Chidambaram]{mohan2020analyzing}
Mohan, J., Phanishayee, A., Raniwala, A., and Chidambaram, V.
\newblock Analyzing and mitigating data stalls in dnn training, 2020.
\newblock URL \url{https://arxiv.org/abs/2007.06775}.

\bibitem[Paszke et~al.(2019)Paszke, Gross, Massa, Lerer, Bradbury, Chanan,
  Killeen, Lin, Gimelshein, Antiga, et~al.]{paszke2019pytorch}
Paszke, A., Gross, S., Massa, F., Lerer, A., Bradbury, J., Chanan, G., Killeen,
  T., Lin, Z., Gimelshein, N., Antiga, L., et~al.
\newblock Pytorch: An imperative style, high-performance deep learning library.
\newblock \emph{Advances in neural information processing systems}, 32, 2019.

\bibitem[{PyTorch Core Team}()]{pytorchwebsite}
{PyTorch Core Team}.
\newblock \emph{{PyTorch}: PyTorch Docs}.
\newblock PyTorch.

\bibitem[Shi et~al.(2016)Shi, Wang, Xu, and Chu]{shi2016benchmarking}
Shi, S., Wang, Q., Xu, P., and Chu, X.
\newblock Benchmarking state-of-the-art deep learning software tools.
\newblock In \emph{2016 7th International Conference on Cloud Computing and Big
  Data (CCBD)}, pp.\  99--104. IEEE, 2016.

\bibitem[Tao et~al.(2018)Tao, Du, Guo, Lan, Zhang, Zhou, Xu, Liu, Liu, Tang,
  et~al.]{tao2018benchip}
Tao, J.-H., Du, Z.-D., Guo, Q., Lan, H.-Y., Zhang, L., Zhou, S.-Y., Xu, L.-J.,
  Liu, C., Liu, H.-F., Tang, S., et~al.
\newblock Benchip: Benchmarking intelligence processors.
\newblock \emph{Journal of Computer Science and Technology}, 33\penalty0
  (1):\penalty0 1--23, 2018.

\bibitem[Team(2022{\natexlab{a}})]{2022ActiveloopHub}
Team, A.~D.
\newblock Hub: A dataset format for ai. a simple api for creating, storing,
  collaborating on ai datasets of any size \& streaming them to ml frameworks
  at scale.
\newblock \emph{GitHub. Note: https://github.com/activeloopai/Hub},
  2022{\natexlab{a}}.

\bibitem[Team(2022{\natexlab{b}})]{2022squirrelcore}
Team, S.~D.
\newblock Squirrel: A python library that enables ml teams to share, load, and
  transform data in a collaborative, flexible, and efficient way.
\newblock \emph{GitHub. Note:
  https://github.com/merantix-momentum/squirrel-core}, 2022{\natexlab{b}}.
\newblock \doi{10.5281/zenodo.6418280}.

\bibitem[TorchData(2021)]{TorchData}
TorchData.
\newblock Torchdata: A prototype library of common modular data loading
  primitives for easily constructing flexible and performant data pipelines.
\newblock \url{https://github.com/pytorch/data}, 2021.

\bibitem[Wang et~al.(2020)Wang, Wei, and Brooks]{wang2020systematic}
Wang, Y., Wei, G.-Y., and Brooks, D.
\newblock A systematic methodology for analysis of deep learning hardware and
  software platforms.
\newblock \emph{Proceedings of Machine Learning and Systems}, 2:\penalty0
  30--43, 2020.

\bibitem[Webdataset(2013)]{Webdataset}
Webdataset.
\newblock Webdataset format.
\newblock \url{https://github.com/webdataset/webdataset}, 2013.

\bibitem[Wu et~al.(2018)Wu, Cao, Sahin, and Liu]{wu2018experimental}
Wu, Y., Cao, W., Sahin, S., and Liu, L.
\newblock Experimental characterizations and analysis of deep learning
  frameworks.
\newblock In \emph{2018 IEEE International Conference on Big Data (Big Data)},
  pp.\  372--377. IEEE, 2018.

\bibitem[Wu et~al.(2019)Wu, Liu, Pu, Cao, Sahin, Wei, and
  Zhang]{wu2019comparative}
Wu, Y., Liu, L., Pu, C., Cao, W., Sahin, S., Wei, W., and Zhang, Q.
\newblock A comparative measurement study of deep learning as a service
  framework.
\newblock \emph{IEEE Transactions on Services Computing}, 2019.

\bibitem[Zhang et~al.(2019)Zhang, Wei, Xu, Jin, and Li]{zhang2019ai}
Zhang, W., Wei, W., Xu, L., Jin, L., and Li, C.
\newblock Ai matrix: A deep learning benchmark for alibaba data centers.
\newblock \emph{arXiv preprint arXiv:1909.10562}, 2019.

\bibitem[Zhu et~al.(2018)Zhu, Akrout, Zheng, Pelegris, Phanishayee, Schroeder,
  and Pekhimenko]{zhu2018tbd}
Zhu, H., Akrout, M., Zheng, B., Pelegris, A., Phanishayee, A., Schroeder, B.,
  and Pekhimenko, G.
\newblock Tbd: Benchmarking and analyzing deep neural network training.
\newblock \emph{arXiv preprint arXiv:1803.06905}, 2018.

\end{thebibliography}
\bibliographystyle{mlsys2022}

\newpage
 \clearpage
\newpage






\appendix

\section{Numerical Results Cont.}
\label{ap:extra-experiments}

In this appendix, we include a collection of plots for which we did not have space in the core pages of the article.

\begin{figure}[h]
    \centering
    \includegraphics{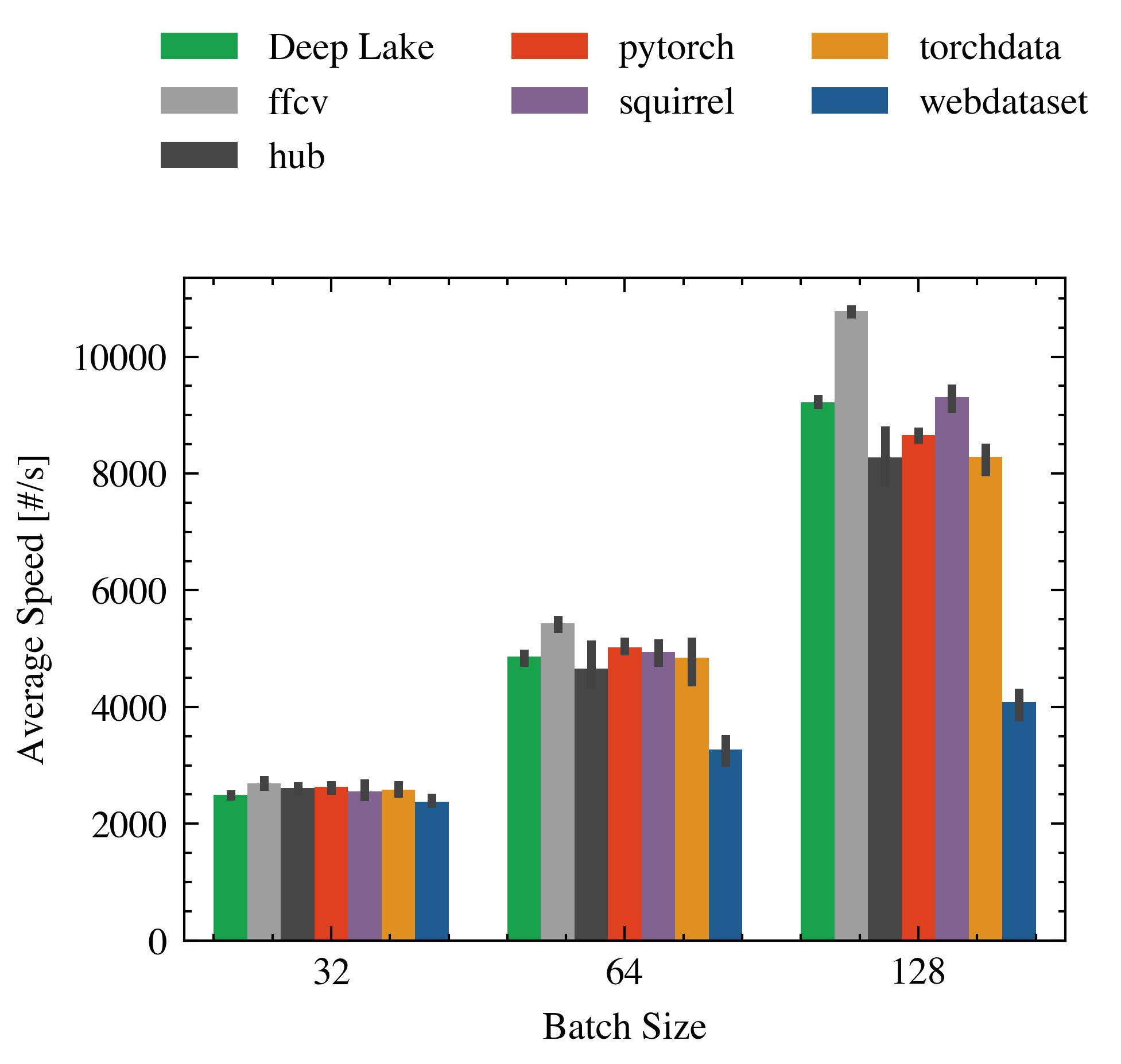}
    \caption{Comparing the impact of batch size in CIFAR10 with a single GPU.}
    \label{fig:cifar-default-batch-1}
\end{figure}

\begin{figure}[H]
    \centering
    \includegraphics{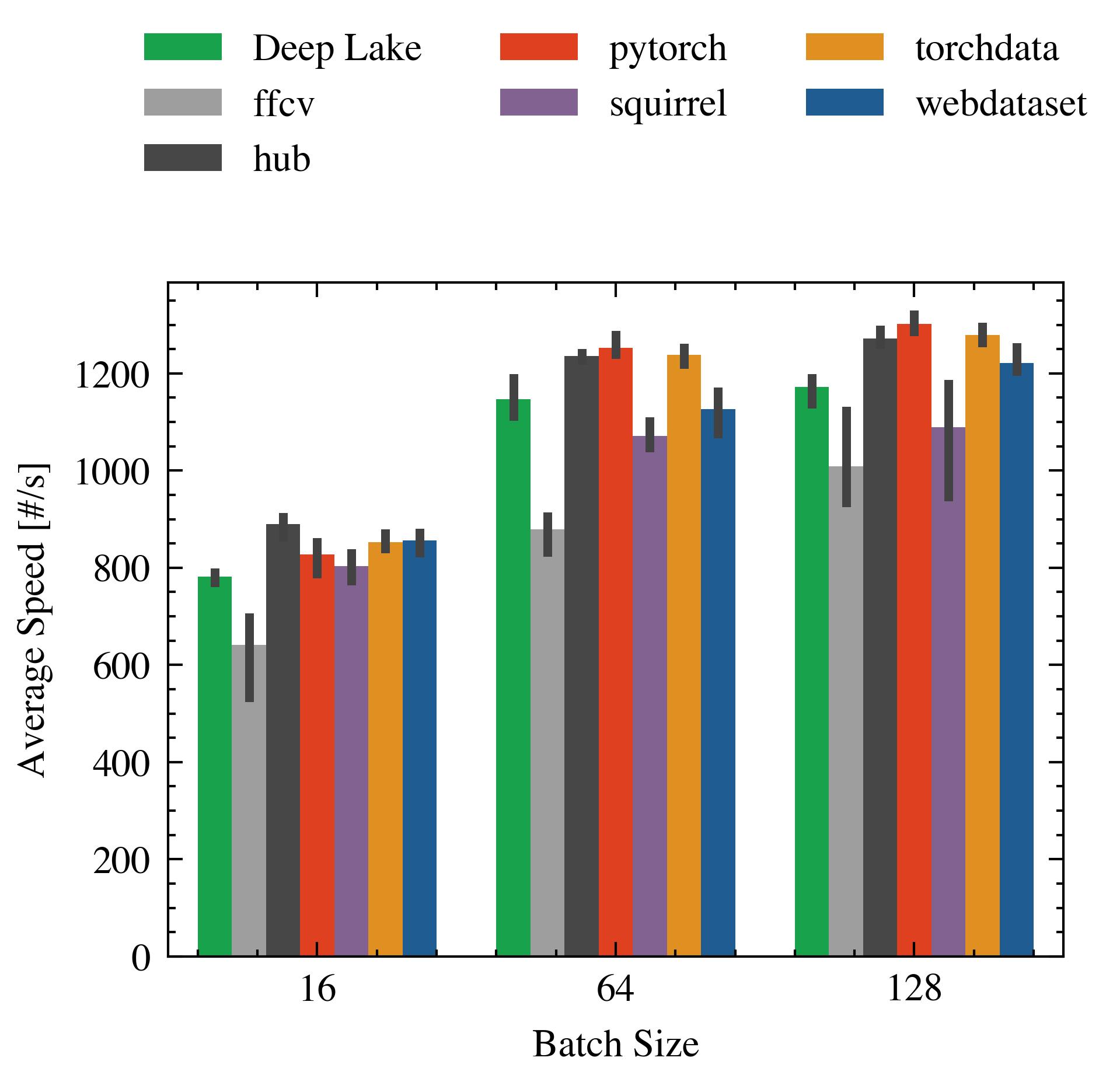}
    \caption{Comparing the impact of batch size in Random with a single GPU.}
    \label{fig:cifar-random-batch-1}
\end{figure}

\begin{figure}[h]
    \centering
    \includegraphics{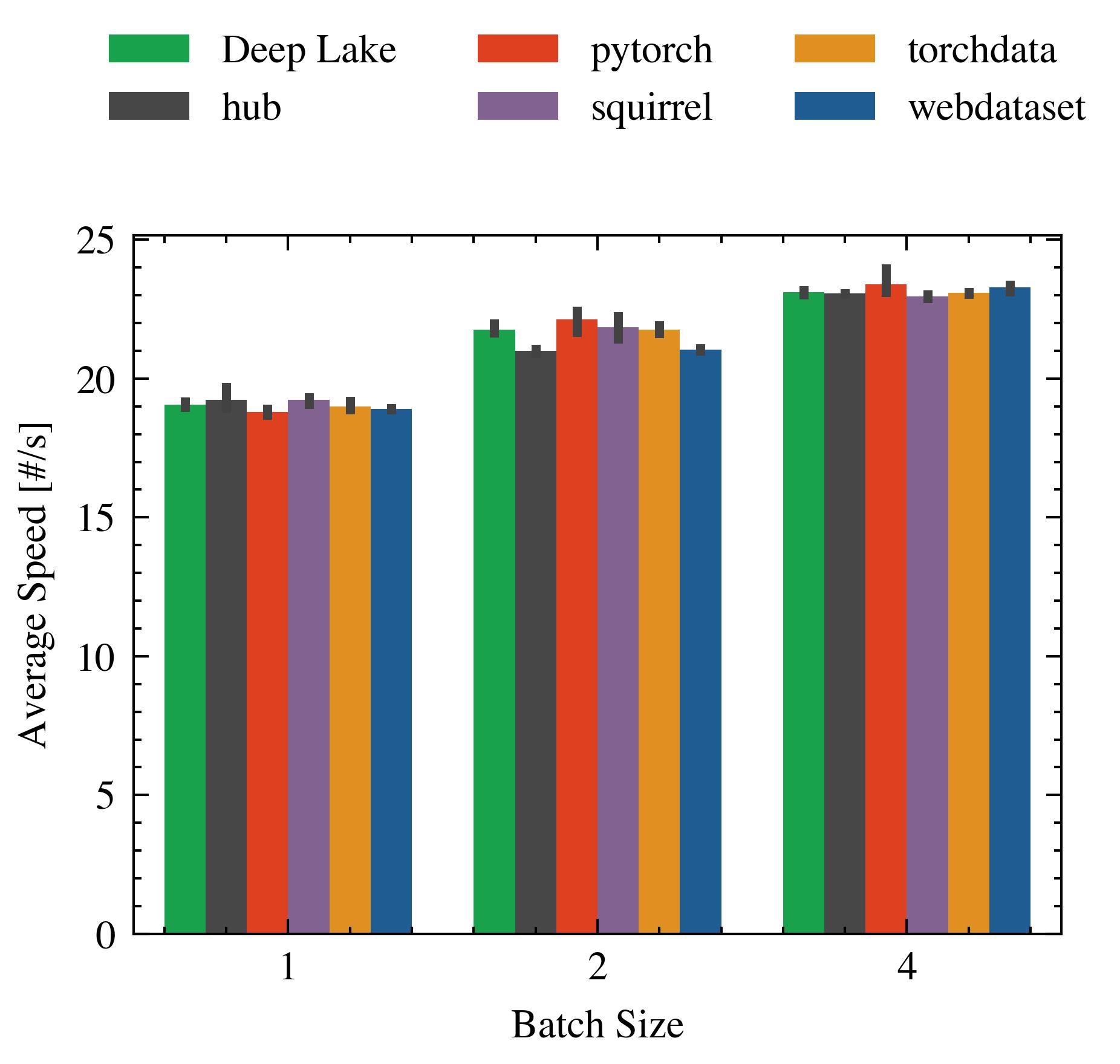}
    \caption{Comparing the impact of batch size in CoCo with a single GPU.}
    \label{fig:cifar-coco-batch-1}
\end{figure}

\begin{figure}[h]
    \centering
    \includegraphics{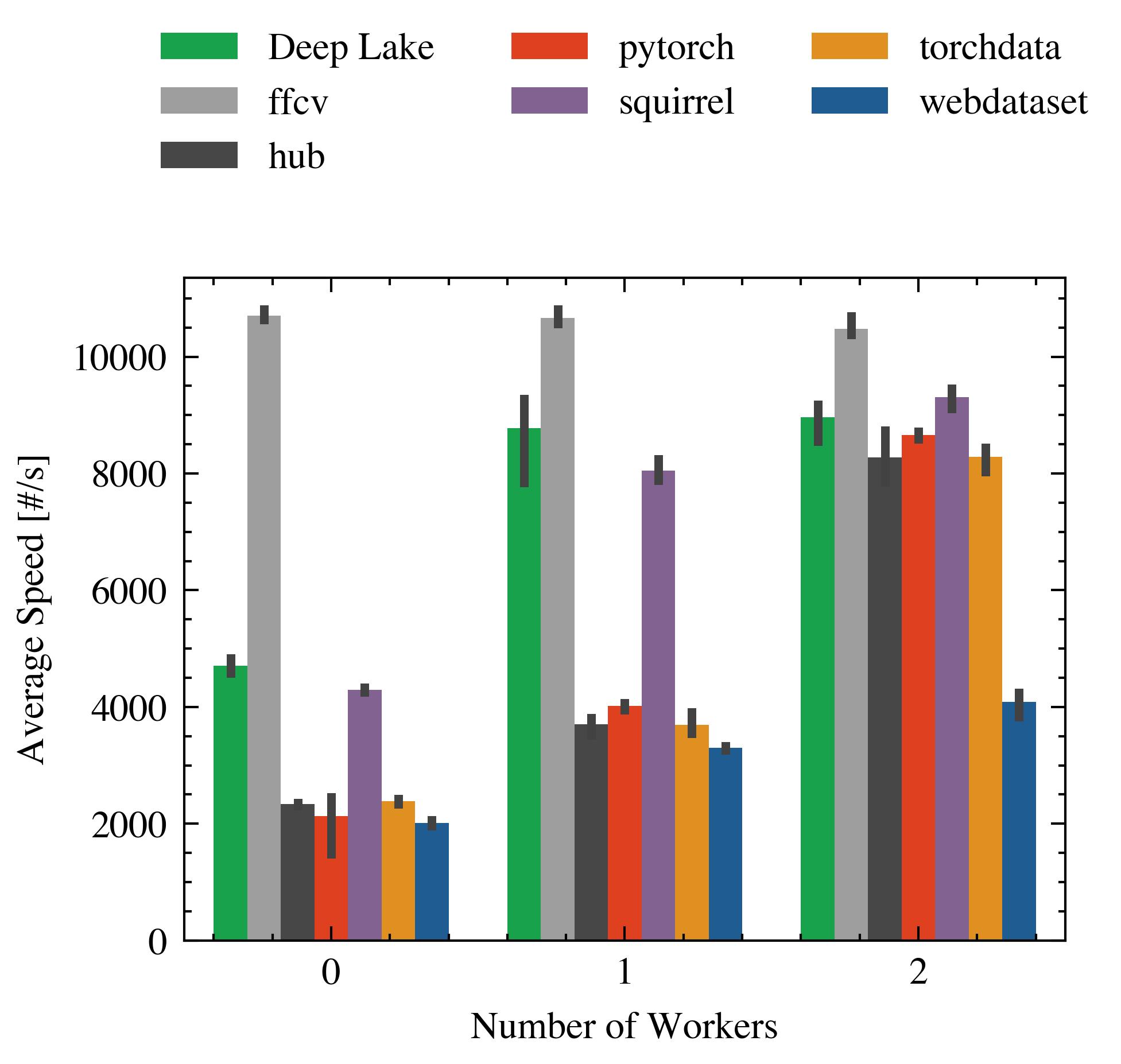}
    \caption{Comparing the impact of the number of workers in CIFAR10 with a single GPU.}
    \label{fig:cifar-default-batch-2}
\end{figure}

\begin{figure}[h]
    \centering
    \includegraphics{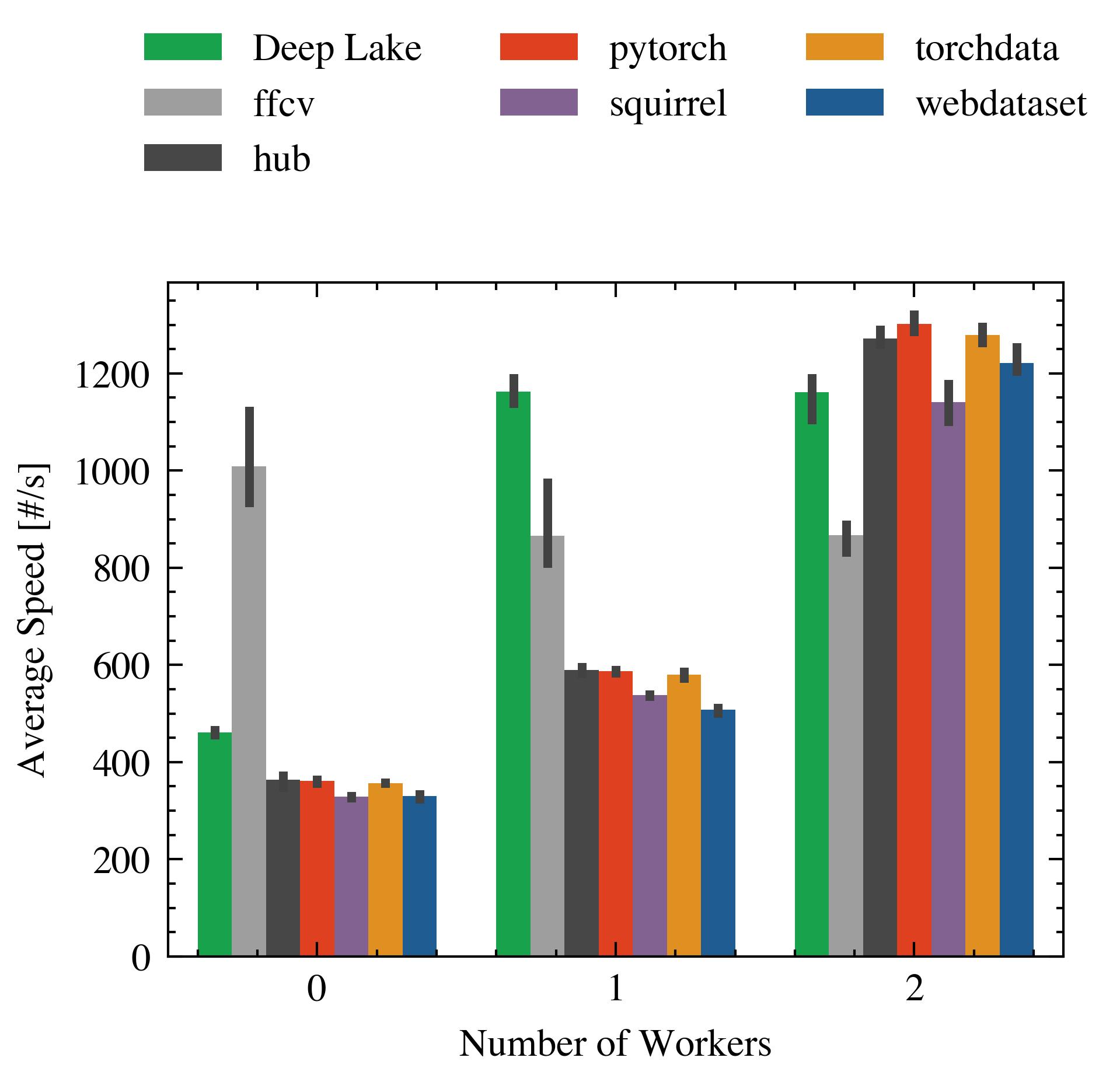}
    \caption{Comparing the impact of the number of workers in Random with a single GPU.}
    \label{fig:cifar-random-batch-2}
\end{figure}

\begin{figure}[h]
    \centering
    \includegraphics{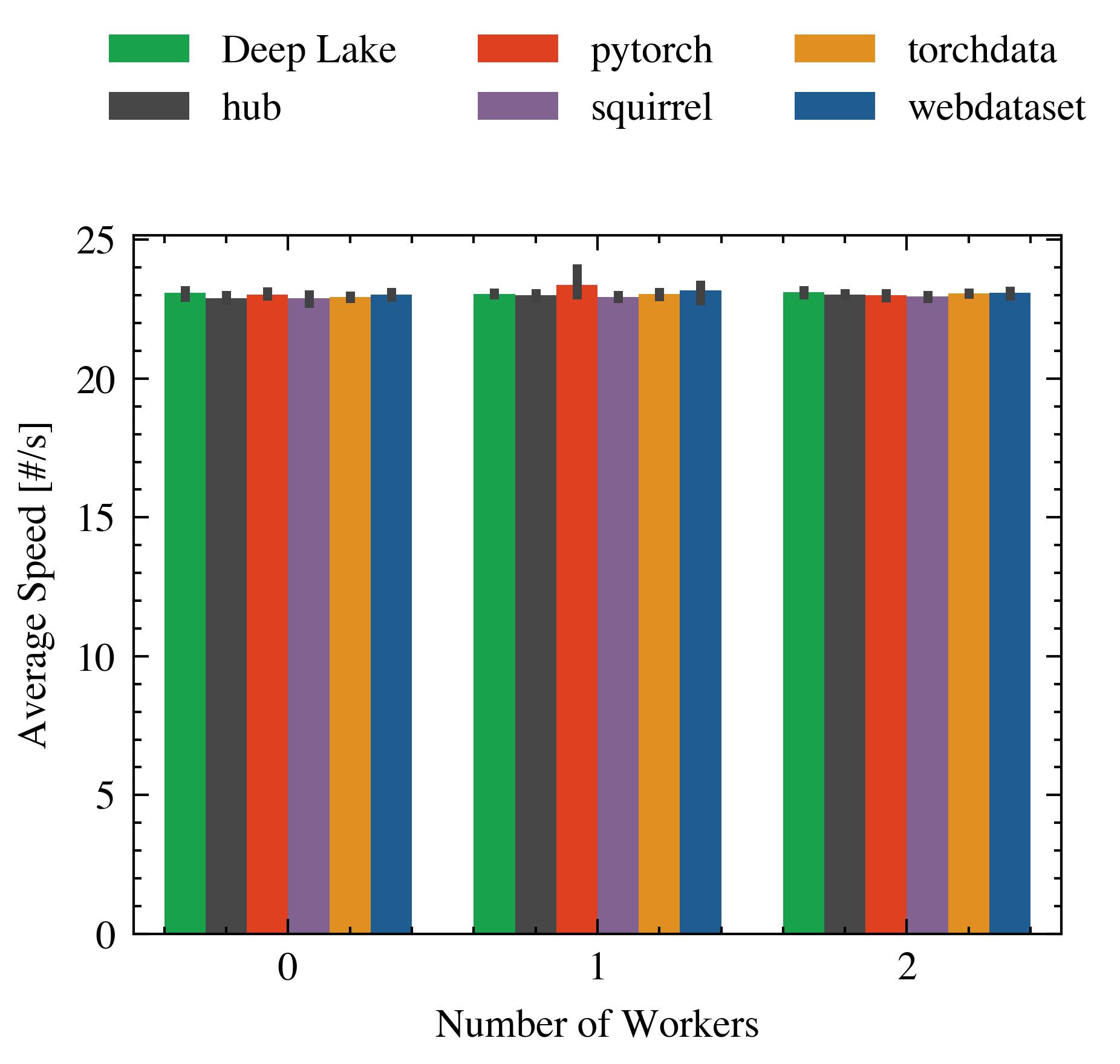}
    \caption{Comparing the impact of the number of workers in CoCo with a single GPU.}
    \label{fig:cifar-coco-batch-2}
\end{figure}


\begin{figure}[h]
    \centering
    \includegraphics{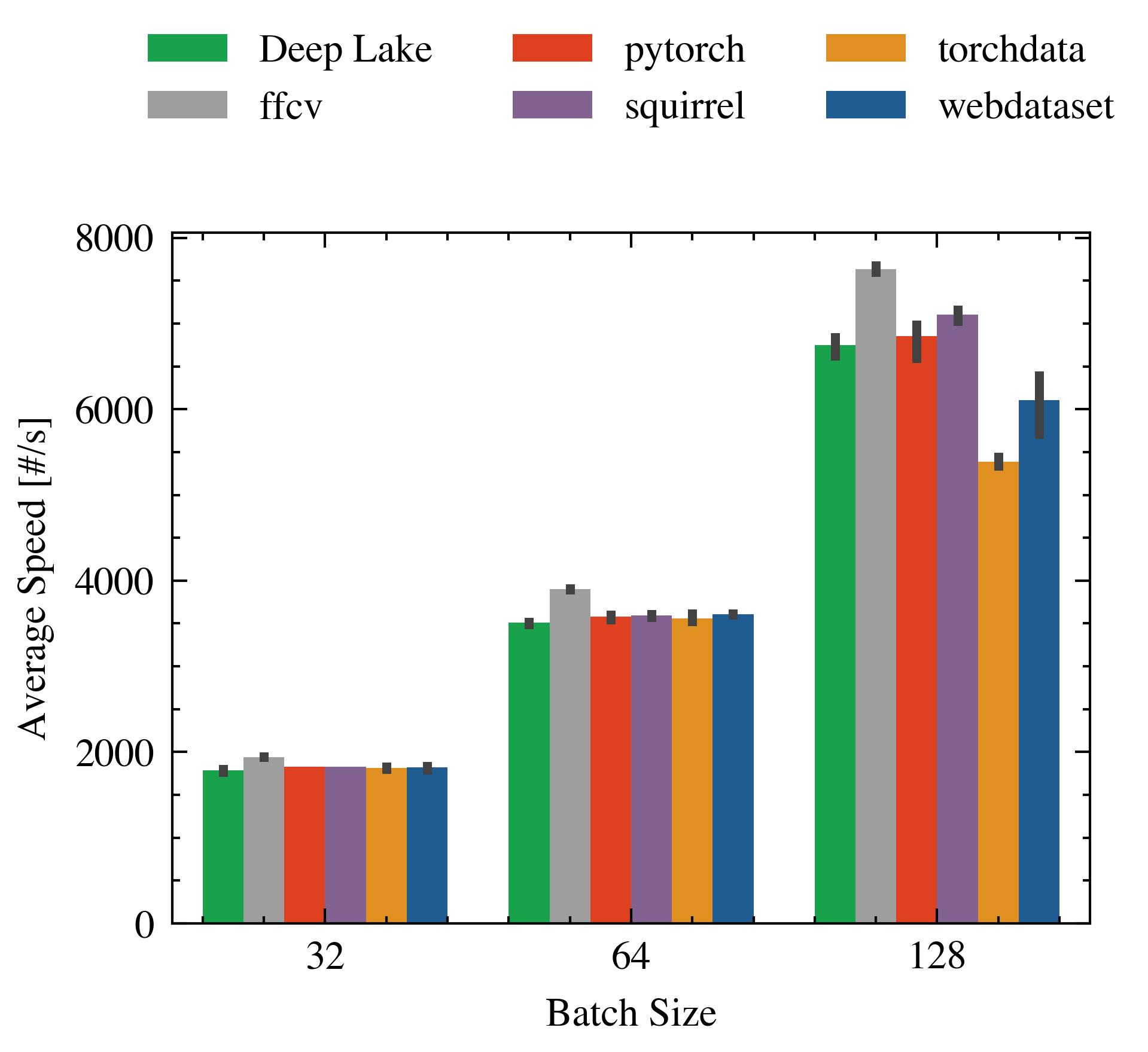}
    \caption{Comparing the impact of batch size in CIFAR10 with multiple GPUs.}
    \label{fig:cifar-default-batch-3}
\end{figure}

\begin{figure}[h]
    \centering
    \includegraphics{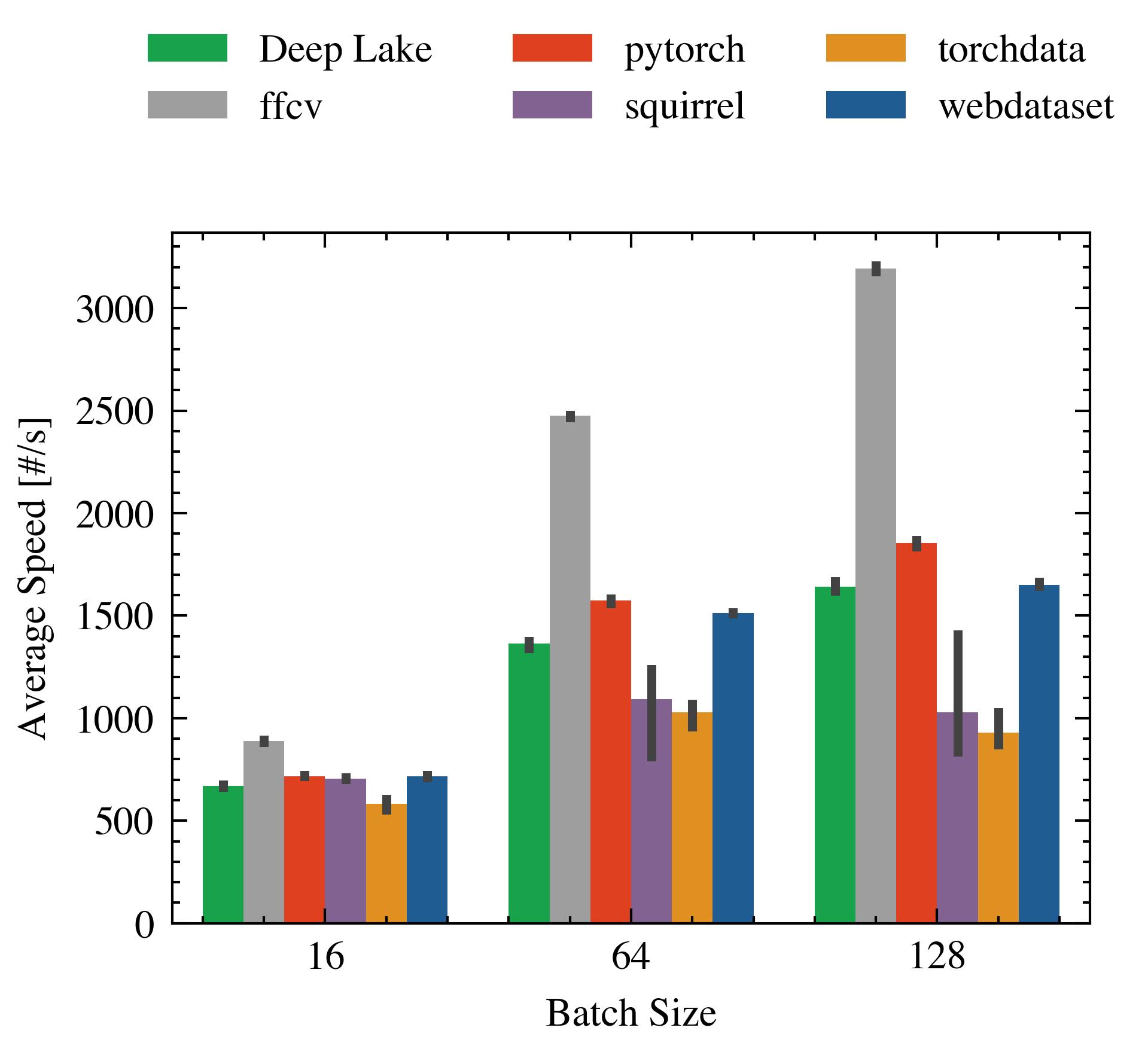}
    \caption{Comparing the impact of batch size in Random with multiple GPUs}
    \label{fig:cifar-random-batch-3}
\end{figure}

\begin{figure}[h]
    \centering
    \includegraphics{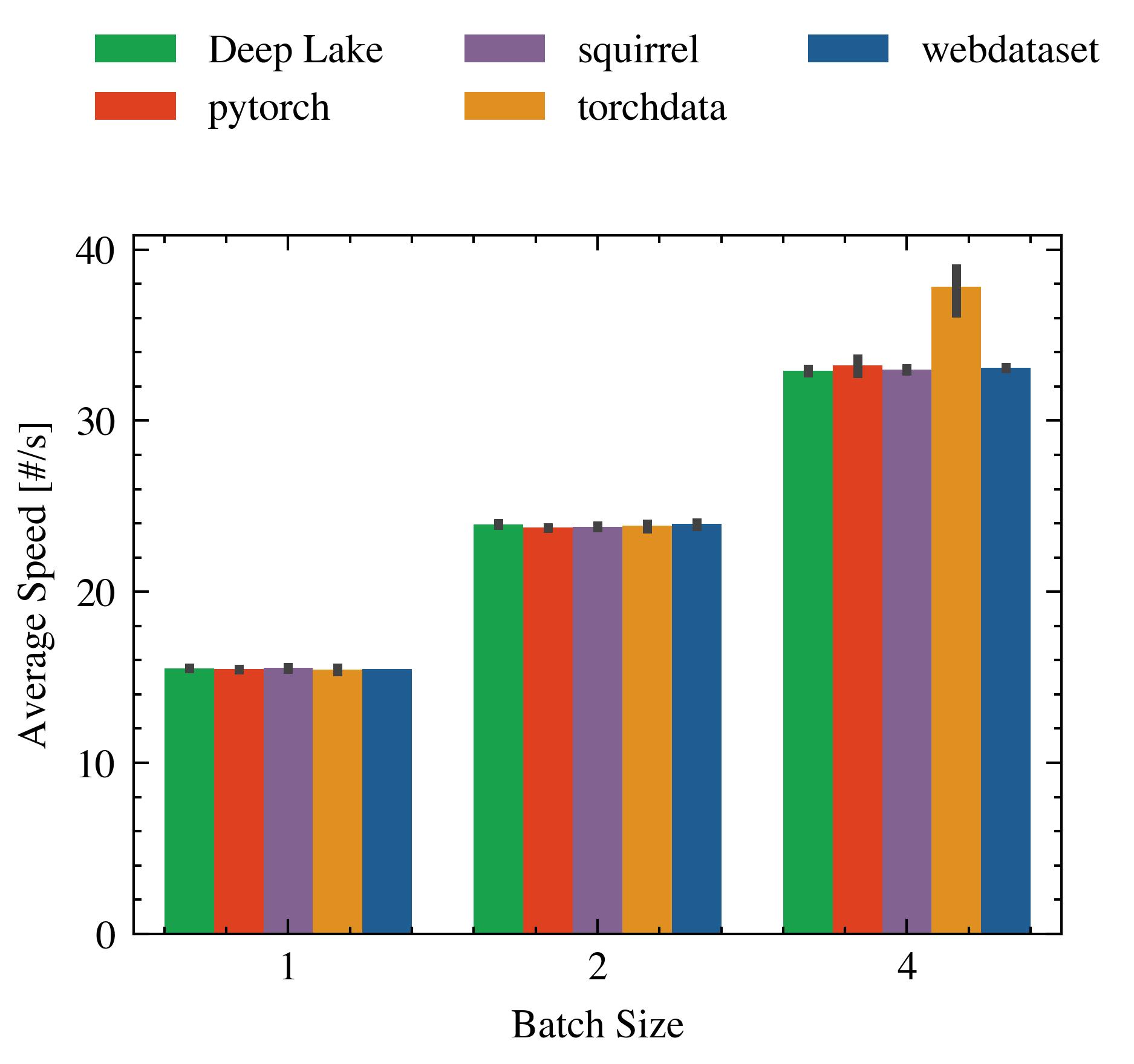}
    \caption{Comparing the impact of batch size in CoCo with multiple GPUs.}
    \label{fig:cifar-coco-batch-3}
\end{figure}

\begin{figure}[h]
    \centering
    \includegraphics{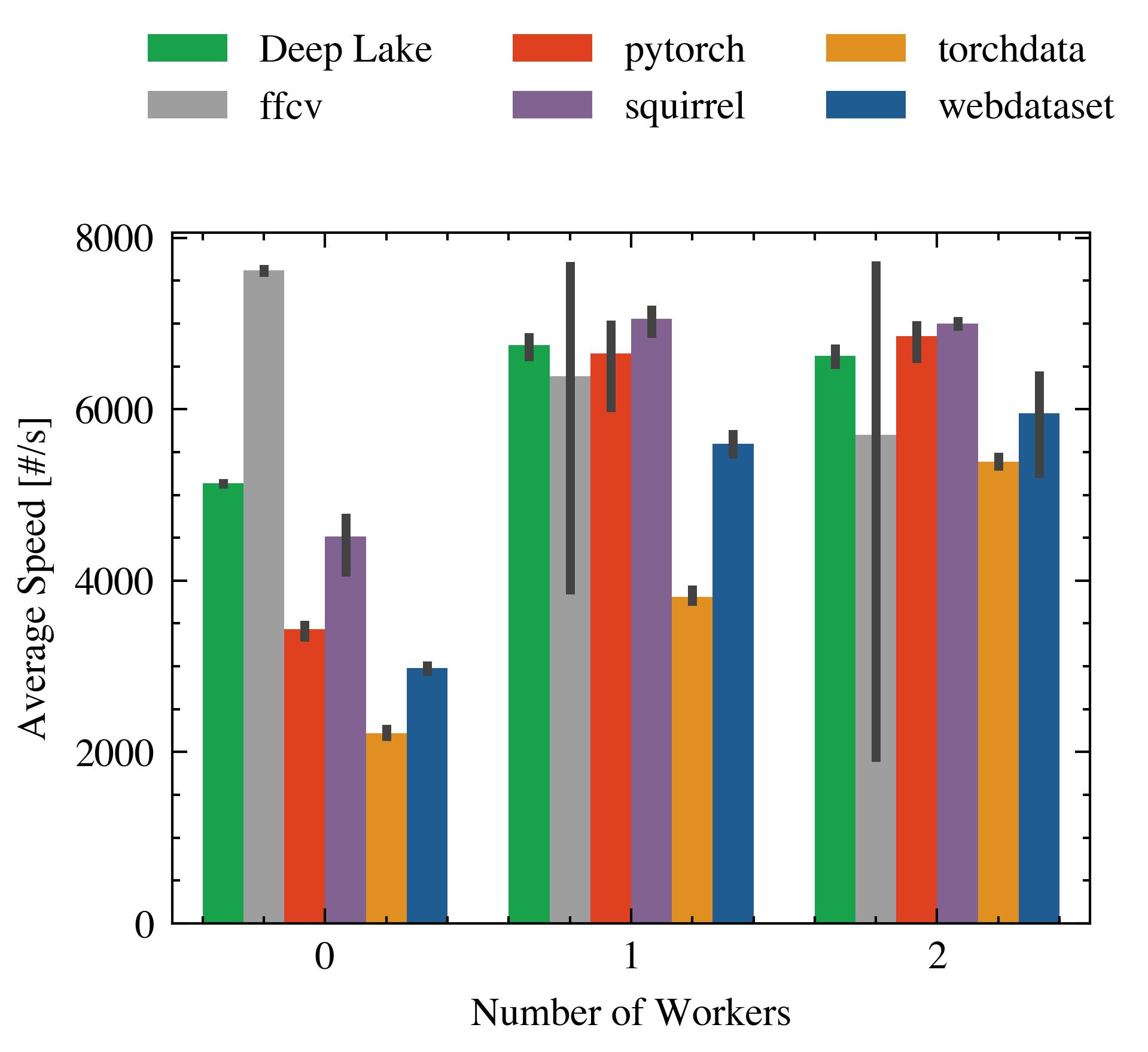}
    \caption{Comparing the impact of the number of workers in CIFAR10 with multiple GPUs}
    \label{fig:cifar-default-batch-4}
\end{figure}

\begin{figure}[h]
    \centering
    \includegraphics{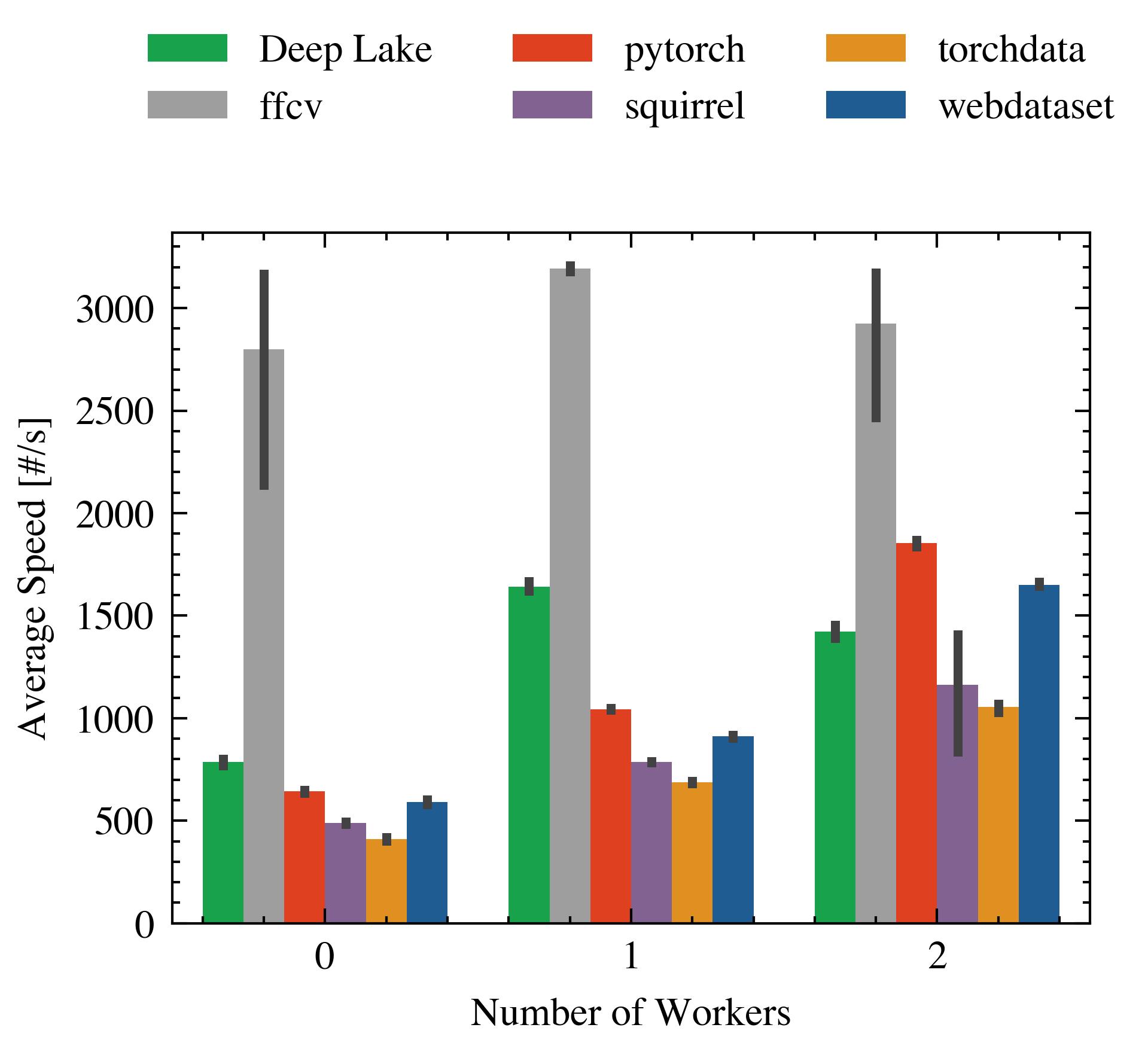}
    \caption{Comparing the impact of the number of workers in Random with multiple GPUs.}
    \label{fig:cifar-random-batch-4}
\end{figure}

\begin{figure}[h]
    \centering
    \includegraphics{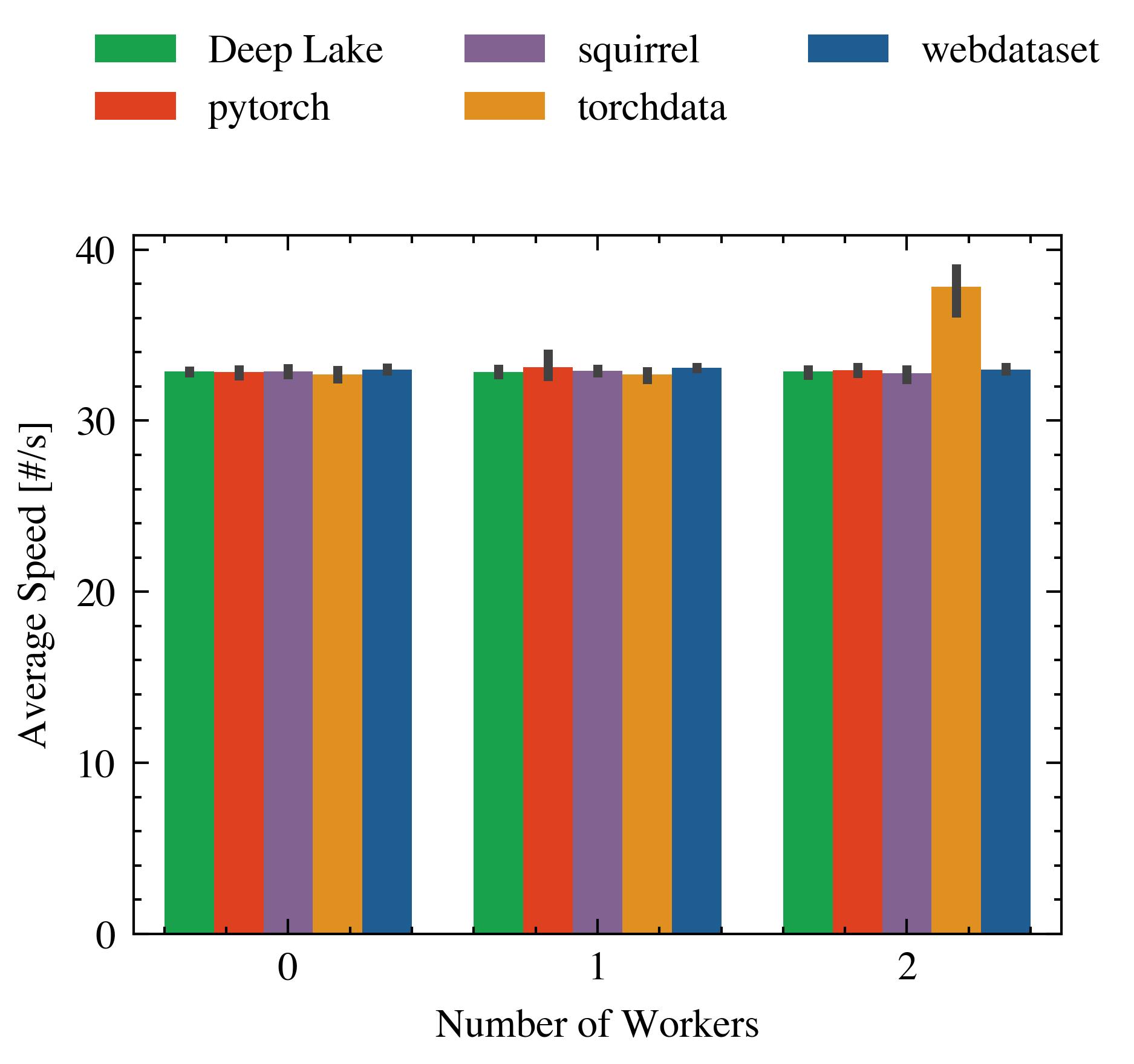}
    \caption{Comparing the impact of the number of workers in CoCo with multiple GPUs.}
    \label{fig:cifar-coco-batch-4}
\end{figure}


\begin{figure}[h]
    \centering
    \includegraphics{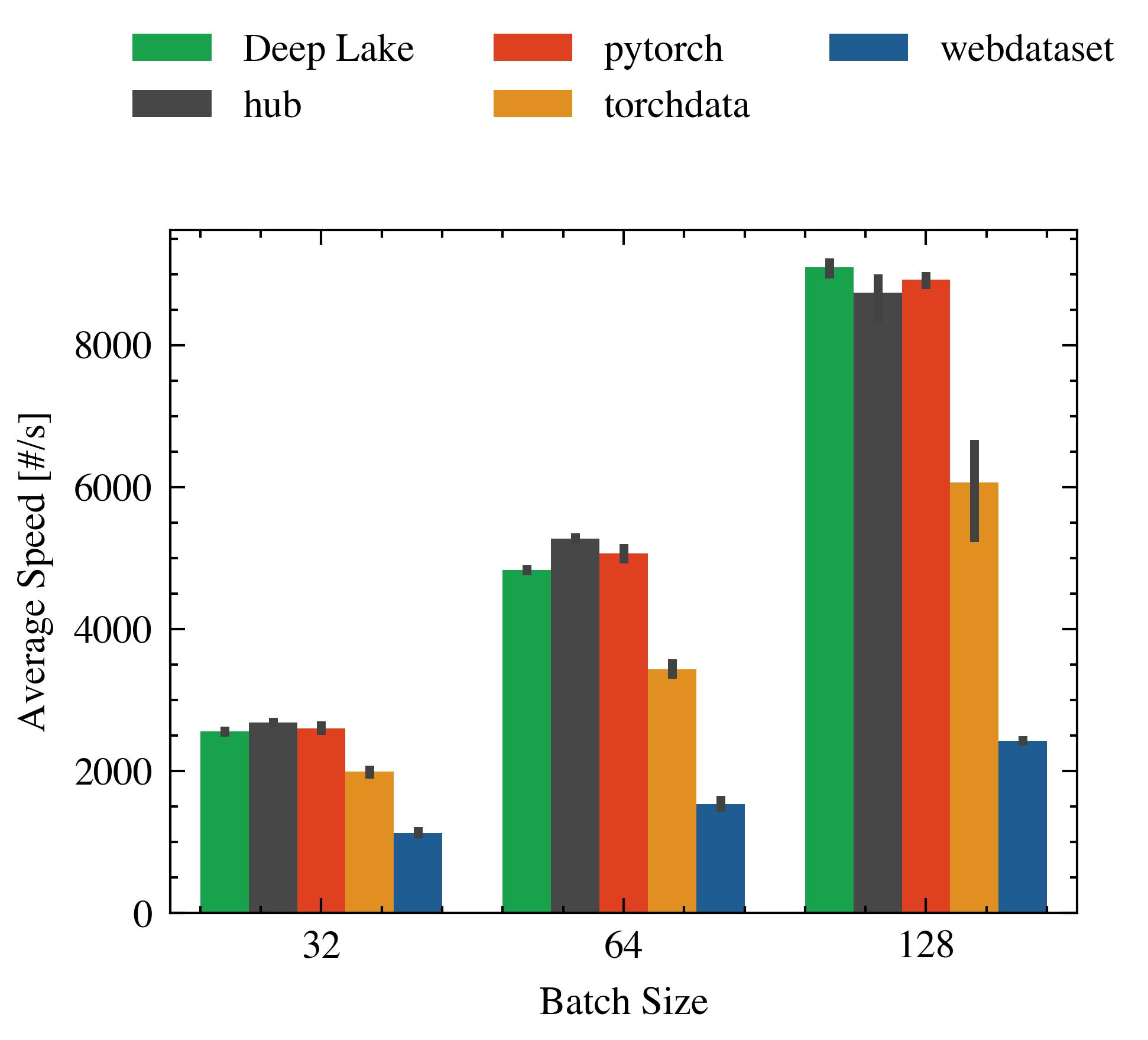}
    \caption{Comparing the impact of batch size in CIFAR10 with a single GPU while filtering.}
    \label{fig:cifar-default-batch-5}
\end{figure}

\begin{figure}[h]
    \centering
    \includegraphics{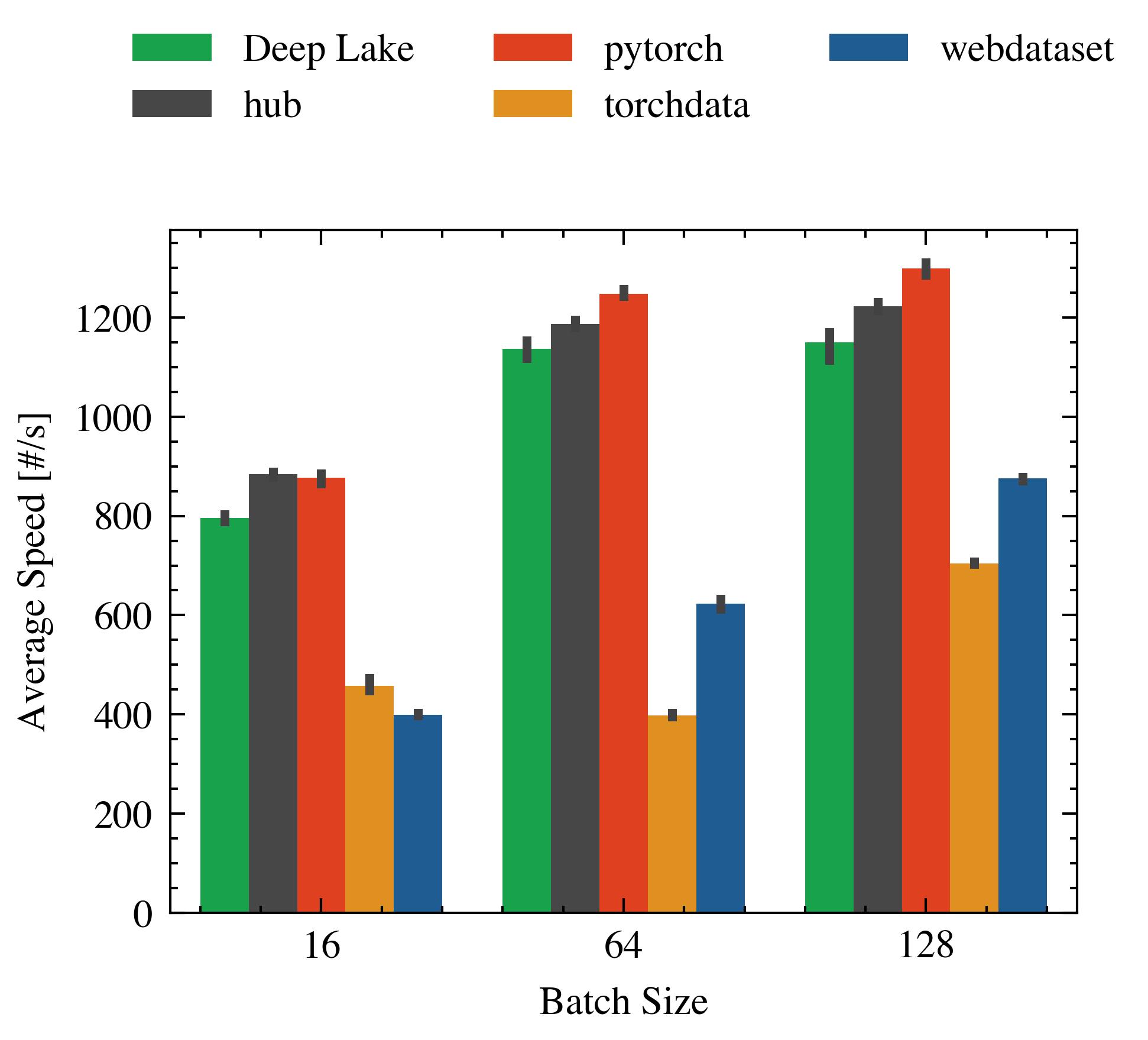}
    \caption{Comparing the impact of batch size in Random with a single GPU while filtering}
    \label{fig:cifar-random-batch-5}
\end{figure}

\begin{figure}[h]
    \centering
    \includegraphics{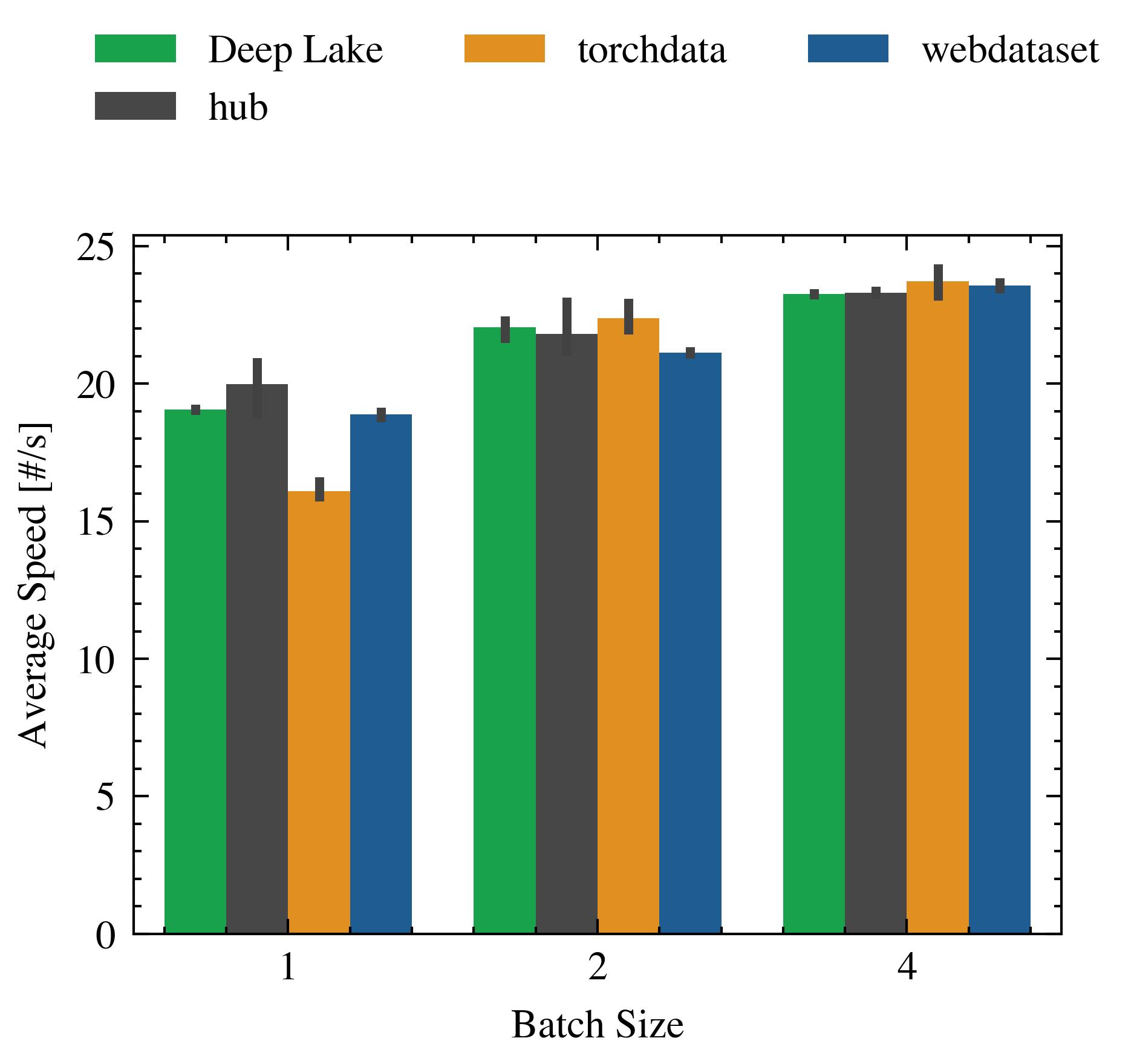}
    \caption{Comparing the impact of batch size in CoCo with a single GPU while filtering.}
    \label{fig:cifar-coco-batch-5}
\end{figure}

\begin{figure}[h]
    \centering
    \includegraphics{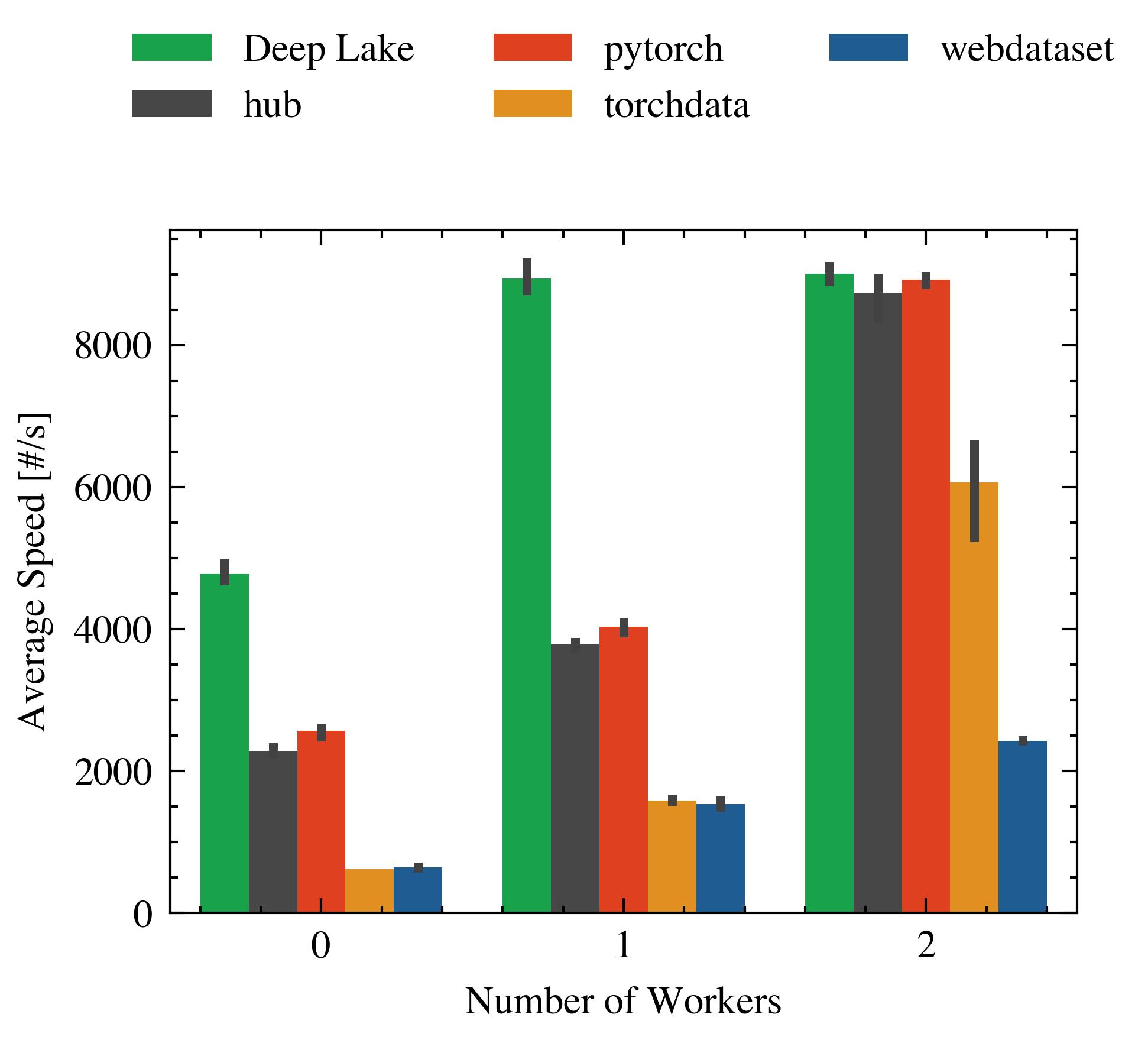}
    \caption{Comparing the impact of the number of workers in CIFAR10 with a single GPU while filtering}
    \label{fig:cifar-default-batch-6}
\end{figure}

\begin{figure}[h]
    \centering
    \includegraphics{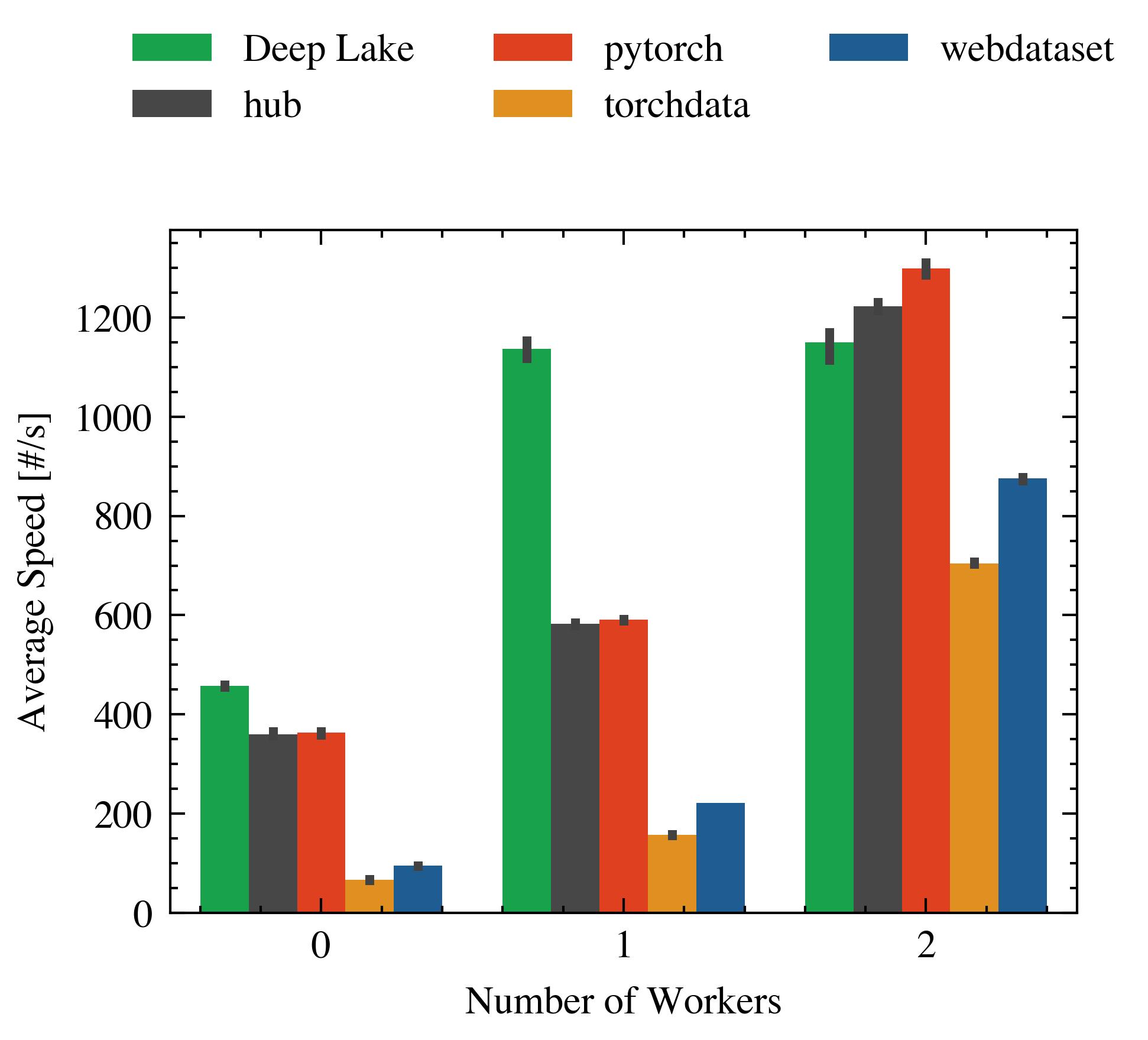}
    \caption{Comparing the impact of the number of workers in Random with a single GPU while filtering.}
    \label{fig:cifar-random-batch-6}
\end{figure}

\begin{figure}[h]
    \centering
    \includegraphics{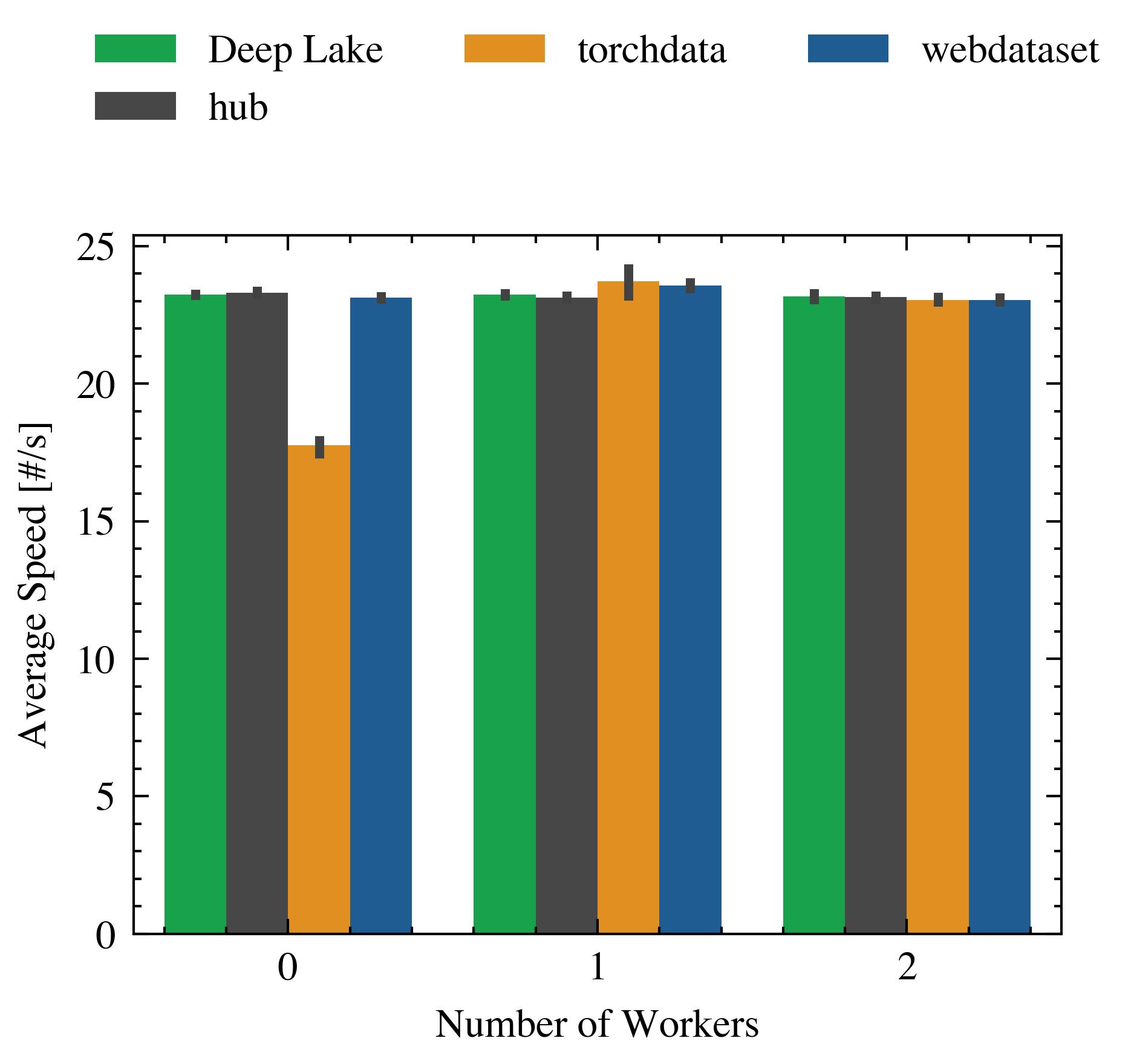}
    \caption{Comparing the impact of the number of workers in CoCo with a single GPU while filtering.}
    \label{fig:cifar-coco-batch-6}
\end{figure}






\end{document}